\newcommand{\bhar}{\ensuremath{\dot{M}_{\bullet}}\xspace}
\newcommand{\athenapk}{\texttt{AthenaPK}\xspace}
\newcommand{\vc}[1]{\textcolor{purple}{#1}}
\newcommand{\cratio}{\ensuremath{\mathcal{C}}\xspace}
\newcommand{\train}{$t_{\mathrm{rain}}$\xspace}
\newcommand{\low}{\texttt{low-turb}\xspace}
\newcommand{\high}{\texttt{high-turb}\xspace}
\newcommand{\stormy}{\emph{stormy}\xspace}
\newcommand{\cloudy}{\emph{cloudy}\xspace}
\newcommand{\rainy}{\emph{rainy}\xspace}
\newcommand{\sunny}{\emph{sunny}\xspace}

\documentclass{aa}  
\usepackage{natbib}
\usepackage{hyperref}
\hypersetup{colorlinks=true,linkcolor=[rgb]{1.,0.2,0.2},citecolor=[rgb]{0.1,0.4,1.},filecolor=[rgb]{0.7,0.2,0.2},urlcolor=[rgb]{0.7,0.2,0.2}}
\usepackage{orcidlink}
\usepackage{graphicx}
\usepackage{txfonts}
\usepackage{booktabs}
\usepackage{amsmath}
\usepackage{amssymb}
\usepackage{verbatim}
\begin{document} 

\title{{\bfseries\scshape BlackHoleWeather –} Jet-regulated chaotic cold accretion across the meso scale:
{\Large Variability and kinematics}}

\titlerunning{Multiphase kinematics and variability diagnostics of jet-regulated CCA}

\authorrunning{V.~Cammelli et al.}

\author{Vieri Cammelli \thanks{vieri.cammelli@unimore.it}
        \inst{1}\orcidlink{0000-0002-2070-9047},
        Massimo Gaspari
        \inst{1}\orcidlink{0000-0003-2754-9258},
        Filippo Barbani
        \inst{1}\orcidlink{0000-0002-1620-2577},
        Olmo Piana
        \inst{1}\orcidlink{0000-0002-1558-5289},
        Giovanni Stel
        \inst{1}\orcidlink{0009-0007-0585-9462},
        Davide M. Brustio
        \inst{1}\orcidlink{0009-0009-7700-1910},
        Valeria Olivares
        \inst{2,3}\orcidlink{0000-0001-6638-4324},
        Roberto Serafinelli
        \inst{4,5}\orcidlink{0000-0003-1200-5071},
        Pasquale Temi
        \inst{6}\orcidlink{0000-0002-8341-342X},
        Ashkbiz Danehkar
        \inst{7}\orcidlink{0000-0003-4552-5997},
        Francesco Salvestrini
        \inst{8}\orcidlink{0000-0003-4751-7421},
        Michael Reefe
        \inst{9}\orcidlink{0000-0003-4701-8497},
        Filippo M. Maccagni
        \inst{10,11}\orcidlink{0000-0002-9930-1844},
        \and
        Francesco Tombesi
        \inst{12, 13, 14}\orcidlink{0000-0002-6562-8654}
        }

\institute{
Department of Physics, Informatics and Mathematics, University of Modena and Reggio Emilia, I-41125 Modena, Italy
\and
Department of Physics, Universidad de Santiago de Chile, Santiago, Chile
\and
CIRAS, Universidad de Santiago de Chile, Santiago, Chile
\and
INAF -- Astronomical Observatory of Rome, 00078 Monte Porzio Catone (Rome), Italy
\and
Instituto de Estudios Astrof\'isicos, Facultad de Ingenier\'ia y Ciencias, Universidad Diego Portales, Avenida Ej\'ercito Libertador 441, Santiago, Chile
\and
NASA Ames Research Center, MS 245-6, Moffett Field, CA 94035-1000, USA
\and
Science and Technology Institute, Universities Space Research Association, Huntsville, AL 35805, USA
\and
INAF -- Osservatorio Astronomico di Trieste, Via G. Tiepolo 11, I-34143 Trieste, Italy
\and
Department of Physics and MIT Kavli Institute for Astrophysics and Space Research, Massachusetts Institute of Technology, Cambridge, MA 02139, USA
\and
INAF -- Osservatorio Astronomico di Cagliari, via della Scienza 5, 09047, Selargius (CA), Italy
\and
Wits Centre for Astrophysics, School of Physics, University of the Witwatersrand, 2000, Johannesburg, South Africa
\and
INAF -- Astronomical Observatory of Rome, 00078 Monte Porzio Catone (Rome), Italy
\and
Department of Physics, University of Rome ``Tor Vergata'', 00133 Rome, Italy
\and
INFN -- Rome ``Tor Vergata'' Section, 00133 Rome, Italy
}


\abstract
{Chaotic cold accretion (CCA) predicts that supermassive black holes are fed by multiphase clouds condensing from turbulent hot atmospheres. In jet-regulated systems, however, cold gas must also remain dynamically connected to the central accretion region.}
{We investigate how a self-regulated kinetic jet modifies the kinematics, radial transport, and variability of CCA across the meso-scale of a typical galaxy-group atmosphere.}
{We develop and analyse two hyper-zoom hydrodynamical simulations of a \(100\) kpc group atmosphere with radiative cooling, driven subsonic turbulence, sub-pc accretion, and a kinetic mass-loaded bipolar jet. The runs differ only in turbulent driving strength. We measure accretion histories, Eddington ratios, power spectra, phase-separated mass fluxes, projected k-plots, and cooling-to-eddy-time (\cratio-ratio) profiles.}
{Both runs become CCA-fed once precipitation begins, with accretion rising from Bondi-like to strongly super-Bondi values while remaining mostly low-Eddington and mechanically dominated. The strongly stirred run develops an early \textit{stormy} phase with extended condensation, bursty feeding, and strong inflow/outflow variability, but later enters a \textit{cloudy} phase in which cold and warm gas persist at meso- and inner macro-scales while sink coupling weakens. The calmer run instead maintains a compact \textit{rainy} state with a longer-lived central reservoir. Accretion-rate spectra show flicker-like low-frequency slopes and red-noise tails; in the \textit{cloudy} phase, the normalization drops and the low-frequency slope flattens. Phase-separated fluxes show fountain-like recycling in the strongly stirred run, but inner-kpc recycling in the calmer run. The jet excavates a hot channel where sustained condensation is suppressed, while \(\cratio\sim1\) is reached most systematically outside the cone and near the jet-ambient interface.}
{Jet-regulated CCA is controlled by meso-scale transport, not only by cold-gas production. Within the {\sc BlackHoleWeather} framework, combined k-plot and \cratio-ratio diagnostics are crucial to distinguish cold gas that is merely present from cold gas dynamically linked to SMBH feeding.}


\maketitle
%

\section{Introduction}
How supermassive black holes (SMBHs) are fed by multiphase gas and how their jets redistribute energy, momentum, and metals through the surrounding atmosphere remain central open problems in galaxy evolution. Within galaxy evolution models, Active Galactic Nuclei (AGNs) are commonly invoked as the drivers of galactic-scale outflows, powered by the release of gravitational energy during gas accretion onto the SMBH at the centre of the potential well \citep[e.g.][]{SilkRees1998, KingPounds2015}. The associated AGN feedback processes, together with stellar feedback from supernovae and stellar winds \citep[e.g.][]{Barbani2023,Barbani2025}, are believed to significantly alter the thermodynamic and dynamical state of the interstellar and circumgalactic media (ISM and CGM), thereby regulating galaxy formation over cosmic time. In physically motivated scenarios, fast winds and relativistic jets launched by AGNs are coupled to the ambient medium \citep[see, e.g.][]{Menci2008, ZubovasKing2012, FaucherGiguere2012}. Nevertheless, the ultimate outcome of this coupling remains unclear, with proposed effects ranging from positive feedback through shock-induced star formation to negative feedback via the suppression of cooling and fragmentation driven by turbulence and radiative heating \citep[][for reviews]{McNamaraNulsen2012,Morganti2017,Mukherjee2018,Gaspari2020}.

Modeling the feedback cycle driven by accretion and outflows in galaxy groups and clusters is particularly challenging because it requires simultaneously capturing physical processes operating across an extreme range of scales. On the largest end, groups and clusters extend over megaparsec distances, whereas the mechanisms responsible for gas accretion onto the SMBH and the subsequent launch of jets and winds are set on sub-parsec scales, with horizon radii $\sim 10^{-5}$\,pc for a $10^{8}\,M_\odot$ black hole. In the immediate vicinity of the SMBH, within the $\sim 10$–$10^2$ Schwarzschild radii, only a minor fraction of the inflowing gas is ultimately accreted, while the majority is re-ejected through feedback \citep[e.g.][]{Gaspari2017}. The gravitational energy released in this region is efficiently transformed into mechanical power, driving either ultrafast nuclear outflows with velocities of the order $0.1c$ or highly collimated relativistic jets, especially in systems hosting rapidly spinning black holes \citep[e.g.][]{Tombesi2010,Fabian2012,GianolliBianchi2024}. These processes are commonly identified as the initial trigger for AGN feedback. 
In parallel, radiation produced in the innermost regions of the accretion disk contributes to an additional feedback channel, although becoming significant only at high Eddington ratios ($\dot{M}/\dot{M}_{\rm Edd} \gtrsim 0.1$) \citep[e.g.][]{Ciotti2010}.

Observationally, AGN perturb their gaseous atmospheres through shocks, bubbles, cavities, uplift, and multiphase outflows, with mechanical powers up to \(\sim 10^{46}\,{\rm erg\,s^{-1}}\) inferred from X-ray cavity and shock measurements \citep{Fabian2012,GittiBrighenti2012,Eckert2021}. Over the last decade, ALMA, MUSE, XMM-Newton, and Chandra have shown that these feedback events are not single-phase phenomena: molecular, warm ionized, and X-ray-emitting gas often coexist over the same central regions, with disturbed velocity fields and broad line profiles \citep{Alatalo2013,Cicone2014,Fiore2017,Mainieri2021,Maiolino2017,Morganti2017,Tombesi2010,Veilleux2020}. Observed outflow rates span roughly \(\dot{M}_{\rm out}\sim0.1\)--\(100\,M_\odot\,{\rm yr^{-1}}\), while characteristic velocities range from \(\sim10^2\) to \(\sim10^4\,{\rm km\,s^{-1}}\), generally decreasing from highly ionized/X-ray gas to the molecular phase \citep{KingPounds2015,TombesiCappi2013}. However, the same observed multiphase structures can trace direct outflow, turbulent stirring, uplift, fallback, or precipitation-driven inflow. Physically grounded kinematic diagnostics are therefore required to distinguish bulk flows from turbulence and to determine whether condensed gas is actually feeding the SMBH or merely circulating through the jet-regulated atmosphere.

A physically motivated framework connecting kinematics, variability, and SMBH feeding is \emph{chaotic cold accretion} (CCA). In this scenario, radiative cooling in a turbulent hot atmosphere drives nonlinear condensation, producing cold clouds and filaments that ``rain'' toward the SMBH when cooling and turbulent mixing operate on comparable timescales \citep{Gaspari2013,Gaspari2015,Gaspari2017}. Collisions among multiphase condensates and chaotic torques can cancel angular momentum, boosting accretion above the smooth hot-mode expectation and producing rapidly variable, flickering SMBH feeding. The resulting jet response is therefore intermittent rather than steady. 
Here we focus on the \emph{meso-scale}, the intermediate $\sim 0.1$--$1$ kpc region linking the unresolved SMBH feeding zone to the larger hot halo. This is the scale where condensation, turbulence, jet-driven uplift, and cloud-cloud interactions determine whether cold gas remains connected to the central sink or is instead stirred and recycled. CCA therefore predicts not only multiphase morphology, but also specific kinematic and temporal signatures across this meso-scale: enhanced velocity dispersion, broadened and multi-component line profiles, coexistence of inflowing, uplifted, and outflowing gas, recurrent feedback cycles, and strong variability in accretion and jet power. A key goal of this paper is to determine which of these signatures remain robust in a jet-regulated atmosphere, and how they can diagnose efficient SMBH feeding versus meso-scale recycling.

On the numerical side, general relativity magneto-hydrodynamic (GRMHD) simulations have made major progress in quantifying near-horizon accretion efficiencies, jet launching, and the dependence of mechanical power on magnetic flux and SMBH spin \citep{Tchekhovskoy2012,Sadowski2017}. Conversely, semi-analytic and idealized galaxy-scale models show that ultrafast nuclear outflows and jets can shock, entrain, and accelerate the ambient medium, producing cooler and broader gas components as energy and momentum are transferred from the nuclear to galactic scales \citep{FaucherGiguere2012,Wagner2013,Zubovas2014,SerafinelliTombesi2019}. The remaining challenge is to bridge these regimes. CCA is intrinsically stochastic: condensation, cloud collisions, angular-momentum cancellation, jet response, uplift, and fallback evolve over many local dynamical times. Capturing this behaviour therefore requires long-duration, high-resolution simulations that follow cooling, turbulence, jet feedback, and multiphase kinematics with sufficient cadence to measure both velocity fields and accretion variability across the meso-scale.


This motivates state-of-the-art computational approaches. Here we leverage the open-access \athenapk\footnote{https://github.com/parthenon-hpc-lab/athenapk} code \citep{Grete2023}, which combines portability, scalability, and astrophysical specialization with GPU-resident memory handling, asynchronous one-sided MPI communication, and block-level tasking. This work is part of the {\sc BlackHoleWeather} project (PI: Gaspari), which aims to build a unified, multiscale description of SMBH feeding and feedback in multiphase atmospheres. The present paper, C26b, is the second of two companion studies based on the same jet-regulated simulation suite. The companion paper, \cite{Cammelli2026a}, hereafter C26a, focuses on morphology and thermodynamics, including phase structure, condensation morphology, cavity evolution, and shock/cocoon phenomenology. Here, we investigate the kinematic and time-domain response of the atmosphere: velocity-field statistics, phase-separated inflow and outflow rates, turbulence-related diagnostics, and the time-resolved coupling between SMBH feeding and jet power. Our goal is to identify which kinematic and variability signatures trace efficient chaotic cold accretion, and which instead indicate that condensed gas is being stirred, uplifted, or recycled across the meso-scale rather than delivered to the central engine. 
A central novelty of this work is therefore to separate multiphase gas formation from multiphase gas delivery: cold gas at meso-scales is not by itself a sufficient proxy for SMBH feeding, unless it remains dynamically coupled to the micro-scale sink.
Detailed synthetic spectroscopic and integral-field diagnostics are deferred to future {\sc BlackHoleWeather} work.

In this paper, we investigate how self-regulated AGN jet feedback reshapes the multiphase gas in a galaxy-group atmosphere. We perform three-dimensional hydrodynamical simulations with \athenapk, resolving the central sink and jet-injection region with sub-pc cell size while following the intragroup medium over tens of kpc. The unresolved disk-to-horizon physics is encoded through an effective accretion and feedback prescription, rather than explicitly resolving the black-hole horizon or the jet-launching region. The simulations include gravity, radiative cooling, externally driven turbulence, sink accretion, and mass-loaded kinetic jet injection. Static mesh refinement (SMR) provides the required dynamic range, allowing us to study how jet-driven feedback couples micro-scale feeding to meso-scale multiphase transport and group-scale gas circulation.

In practice, we perform consistent three-dimensional hydrodynamic (HD) simulations targeting galaxy-group environments, spanning from the inner launching region (resolved to sub-parsec scales via static mesh refinement, SMR) up to group/cluster scales (tens of kpc). We follow multiphase gas evolution under gravity, cooling, turbulence driving, jet launch and energy deposition within \athenapk, and extract kinematic and time-domain diagnostics relevant to CCA and feedback self-regulation.

Within the {\sc BlackHoleWeather} series, the present work is also tied to two complementary sets of companion simulations that provide useful reference points for later comparisons. The no-jet CCA simulations of \cite{Barbani2026a, Barbani2026b}, hereafter B26a,b, isolate how ambient turbulence regulates multiphase condensation, radial transport, and pure feeding/rain in the absence of self-regulated jet feedback, thereby providing the natural baseline against which we assess what the jet changes. The spin-regulated simulations of \cite{Piana2026a, Piana2026b}, hereafter P26a,b, instead explore how unresolved angular-momentum transport, SMBH spin evolution, and relativistic prescriptions affect the innermost feeding-feedback loop.

The paper is organized as follows. \S\ref{sec:methods} summarizes the subgrid accretion and jet-feedback prescription that links sink-scale gas inflow to SMBH feeding and mechanical outflow, and  describes the numerical framework, initial conditions, and reference simulations. \S\ref{sec:results} presents the main results on accretion variability, radial inflow/outflow rates, projected gas kinematics, and CCA diagnostics. \S\ref{sec:disc} discusses the implications for jet-regulated CCA, observational interpretation, and AGN self-regulation across different weather states. Finally, \S\ref{sec:concl} summarizes our conclusions.

\section{Numerical methods}\label{sec:methods}

Our simulations are performed with \athenapk, the GPU-accelerated extension of \texttt{Athena++}, built on the Parthenon and \texttt{Kokkos} frameworks \citep{Stone2020,Grete2023,Edwards2014}. Since the full numerical setup is presented in C26a and B26a, here we summarize only the ingredients most relevant for the present analysis.

We simulate a $100$ kpc box with static mesh refinement (SMR) over 10 levels, starting from a $128^3$ root grid and reaching a finest spatial resolution of $\Delta x_{\min}\simeq 0.78$ pc. The inner sink region is centred at the origin and has radius $r_{\rm sink}=4\Delta x_{\min}$ (cf.~\citealt{Gaspari2013} and B26a). We adopt the same gravitational potential, radiative cooling, turbulence driving, and initial conditions as in C26a and B26a. Magnetic fields are neglected in this work in order to isolate the role of kinetic feedback and ambient turbulence.

The gas dynamics are evolved in conservative Eulerian form as
\begin{equation}
\frac{\partial \rho}{\partial t} + \nabla\cdot(\rho\boldsymbol{v}) = \mathcal{S}^{\rm jet}_{\rho},
\end{equation}
\begin{equation}
\frac{\partial (\rho\boldsymbol{v})}{\partial t} + \nabla\cdot(\rho\boldsymbol{v}\otimes\boldsymbol{v}+P\mathbb{I}) = -\rho\nabla\Phi + \rho\boldsymbol{f}_{\rm turb} + \boldsymbol{\mathcal{S}}^{\rm jet}_{\rho\boldsymbol{v}},
\end{equation}
\begin{equation}
\frac{\partial E}{\partial t} + \nabla\cdot\big[(E+P)\boldsymbol{v}\big] =  -\rho\boldsymbol{v}\cdot\nabla\Phi - \mathcal{L} +\mathcal{S}_{\rm turb} +\mathcal{S}^{\rm jet}_{E},
\end{equation}
where $\mathcal{L}=n^2\Lambda(T)$ is the radiative cooling term and $\mathcal{S}_{\rm turb}=\rho\boldsymbol{v}\cdot\boldsymbol{f}_{\rm turb}$ is the turbulence-driving source term. We use the same second-order Godunov scheme, PLM reconstruction, HLLC solver, and fallback strategy described in C26a.

\subsection{Subgrid feedback prescription}\label{subsec:subgrid}

Because the sink particle does not resolve the flow from the accretion disc down to the black-hole horizon or the innermost stable circular orbit (ISCO), we map the sink accretion rate, $\dot{M}_{\rm sink}$, to an SMBH accretion rate, $\dot{M}_{\bullet}$, through
\begin{equation}\label{eq:bhar}
    \dot{M}_{\bullet} = \big(1-\epsilon_{\rm disk}\big)\dot{M}_{\rm sink},
\end{equation}
where $\epsilon_{\rm disk}$ denotes the fraction of gravitational energy dissipated between the sink radius and the ISCO. Under the assumption of a radiatively efficient disc, this is equivalent to the disc radiative efficiency. In principle these efficiencies depend on BH spin; here we keep them fixed to isolate the impact of feedback (see P26a for more details on spin-jet coupling).

The unresolved SMBH feeding rate is coupled, neglecting magnetic fields, to the injected feedback power through
\begin{equation}\label{eq:tot_power}
    P_{\rm tot} = P_{\rm kin}+P_{\rm th}
    = \epsilon_{\rm tot}\dot{M}_{\bullet}c^2
    \equiv \big(\epsilon_{\rm kin}+\epsilon_{\rm th}\big)\dot{M}_{\bullet}c^2,
\end{equation}
with the corresponding mass-loaded jet rate
\begin{equation}\label{eq:mdot_out}
    \dot{M}_{\rm jet} \equiv \eta_{\rm jet}\dot{M}_{\bullet},
\end{equation}
where \(\eta_{\rm jet}\) is the adopted mass-loading factor (B26a). In these fiducial runs we set \(\eta_{\rm jet}=1-\epsilon_{\rm tot}\), so that most of the gas reaching the unresolved nuclear region is returned to the resolved atmosphere through the jet. This choice is motivated by GRMHD models, in which only a small fraction of the inflowing gas is ultimately accreted while the rest is expelled in mechanically dominated outflows \citep[e.g.][]{Sadowski2017,GaspariSadowski2017}.

In this work, we consider a kinetically-dominated jet, and retain the thermal term only as an effective internal-energy contribution at injection. In this limit
\begin{equation}\label{eq:approx_tot_power}
    P_{\rm tot} \simeq P_{\rm kin}+P_{\rm th,jet}
    = \epsilon_{\rm tot}\dot{M}_{\bullet}c^2,
\end{equation}
where
\begin{equation}\label{eq:kin_power}
    P_{\rm kin} = \frac{1}{2}\dot{M}_{\rm jet}v_{\rm jet}^2,
\end{equation}
\begin{equation}\label{eq:th_power}
    P_{\rm th,jet} = e_{\rm jet}\,\dot{M}_{\rm jet},
\end{equation}
and $e_{\rm jet} = {k_{\rm B} T_{\rm jet}}/[(\gamma-1)(\mu m_{\rm p})]$ is the internal energy per unit mass of the injected gas. Combining the above equations yields
\begin{equation}\label{eq:v_out}
    v_{\rm jet} =
    \sqrt{\frac{2}{1-\epsilon_{\rm tot}}
    \Big[\epsilon_{\rm tot}c^2-(1-\epsilon_{\rm tot}) e_{\rm jet}\Big]}.
\end{equation}

This prescription should be interpreted as a controlled subgrid coupling between the pc-scale sink and the resolved atmosphere: it does not attempt to model the disc-to-horizon launching process directly, but imposes a fixed relation between unresolved feeding and injected mechanical power.

\subsection{Jet injection}\label{subsec:jet}

The feedback module injects mass, momentum, and energy in two oppositely directed cylindrical regions centred on the sink and aligned with the $\pm z$ axis. The jet geometry and numerical implementation follow C26a. Within the injection region, the source terms are
\begin{align}
    \mathcal{S}^{\rm jet}_{\rho} &=
    \dot{\rho}_{\rm jet}(W_+ + W_-),\\
    \boldsymbol{\mathcal{S}}^{\rm jet}_{\rho\boldsymbol{v}} &=
    \dot{\rho}_{\rm jet}v_z\hat{\boldsymbol{z}}\,W_+
    - \dot{\rho}_{\rm jet}v_z\hat{\boldsymbol{z}}\,W_-,\\
    \mathcal{S}^{\rm jet}_{E} &=
    \dot{\rho}_{\rm jet}
    \left(e_{\rm jet}+\frac{v_z^2}{2}\right)(W_+ + W_-),
\end{align}
where $v_z=v_{\rm jet}$ is given by Eq.~(\ref{eq:v_out}), $\dot{\rho}_{\rm jet}$ is set by Eq.~(\ref{eq:mdot_out}) and the injection volume, and $W_{\pm}$ are normalized kernels describing the two jet lobes. As in C26a, the jet is injected with a fixed axis and no opening angle, allowing us to isolate the role of ambient turbulence in shaping the resolved kinematics and variability.

\subsection{Reference runs}\label{subsec:ref_runs}

We analyze the same two reference runs presented in C26a, which differ only in the level of initial turbulence driving. They bracket two atmospheric ``weather'' regimes: a \texttt{low-turb} case with weaker non-thermal support and earlier, more centrally retained condensation, and a \texttt{high-turb} case with stronger stirring, delayed precipitation, and more extended multiphase development. In both runs, turbulence is first allowed to develop before radiative cooling and jet feedback are activated; the subsequent evolution is analyzed in terms of the rain time $t_{\rm rain}$ defined in C26a.

\subsection{Initial conditions \& simulation suite}\label{subsec:sim_suite}
\begin{table}
\caption{Summary of the numerical setup and reference runs. Parameters common to both runs are listed once; run-dependent quantities are given in the last two columns. The acceleration field has an RMS amplitude $a_{\mathrm{rms}}$ and is injected with a solenoidal fraction $\zeta=1$, corresponding to purely incompressible driving peaking at mode number $n_{\mathrm{peak,v}}$. The $N_{\mathrm{modes,v}}$ random Fourier modes refresh every correlation time $t_{\mathrm{corr}}$.}
\label{tab:setup}
\centering
\begin{tabular}{lcc}
\hline
Parameter & \low & \high \\
\hline
\multicolumn{3}{c}{\textit{Common numerical setup}} \\
\hline
$L_{\rm box}\ [{\rm kpc}]$ & \multicolumn{2}{c}{$100$} \\
Root grid & \multicolumn{2}{c}{$128^3$} \\
SMR levels & \multicolumn{2}{c}{$10$} \\
$\Delta x_{\min}\ [{\rm pc}]$ & \multicolumn{2}{c}{$0.78$} \\
$r_{\rm sink}\ [{\rm pc}]$ & \multicolumn{2}{c}{$\simeq 3.1$} \\
$T_{\rm sink}\ [K]$ & \multicolumn{2}{c}{$1\ {\rm K}$} \\
\(\rho_{\rm sink}\) [g cm\(^{-3}\)] & \multicolumn{2}{c}{\(1\times10^{-30}\)} \\
$d_{\rm jet}\ [{\rm pc}]$ & \multicolumn{2}{c}{$\simeq 1.5$} \\
$M_{\bullet}\ [M_\odot]$ & \multicolumn{2}{c}{$2.8\times10^{8}$} \\
$\epsilon_{\rm jet}(=\epsilon_{\rm tot})$ & \multicolumn{2}{c}{$0.028$} \\
$T_{\rm jet}\ [K]$ & \multicolumn{2}{c}{$5\times10^{8}\ {\rm K}$} \\
$v_{\rm jet}$ & \multicolumn{2}{c}{$\simeq0.25\ c$} \\
$n_{\rm peak,v}$ & \multicolumn{2}{c}{$4$} \\
$l_{\rm inj}\ [\rm kpc]$ & \multicolumn{2}{c}{$L_{\rm box}/n_{\rm peak}$} \\
$N_{\rm modes,v}$ & \multicolumn{2}{c}{$64$} \\
$t_{\rm corr}\ [{\rm Myr}]$ & \multicolumn{2}{c}{$30$} \\
\hline
\multicolumn{3}{c}{\textit{Run-dependent quantities}} \\
\hline
$a_{\rm rms}\ [10^{-8}\ {\rm cm\,s^{-2}}]$ & $0.62$ & $1.55$ \\
3D Mach number, $\mathcal{M}$ & $\sim 0.15$ & $\sim 0.4$ \\
$\sigma_v (t_{\rm restart})\ [\rm km\,s^{-1}]$ & $60$--$90$ & $210$--$230$ \\
$t_{\rm rain}\ [\rm Myr]$ & $\sim 9$ & $\sim 16$ \\
Total duration after restart $\ [\rm Myr]$ & $\sim 65$ & $\sim 115$ \\
Total simulated duration $\ [\rm Myr]$ & $\sim 100$ & $\sim 150$ \\
\hline
\end{tabular}
\end{table}

As introduced above, all simulations carried out in this work have been performed with the \athenapk code using a $100$ kpc side box at 10 static refinements levels, which aligns with the resolution of C26a. Throughout the whole set of simulations, we assume the same initial density and temperature profiles as for a static hot gaseous halo in hydrostatic equilibrium. 
The static gravitational potential includes the dark matter halo, the central dominant group galaxy, and the SMBH, whose mass is updated according to Eq.~\ref{eq:bhar}. The initial profiles are discussed in detail in B26a, including the data sample and the fitting procedure used to obtain the initial profiles. Here we note that we opt to simulate a galaxy group (reference parameter values as in B26a) as this configuration shows a consistently faster evolution with respect to a galaxy cluster. In fact, the cooling time in a galaxy group, which drives the criterion for the condensation of gas into CCA episodes \citep{Gaspari2013, Gaspari2017}, results in about $t_{\text{cool}}\approx 10-20$ Myr in the central kpc in our simulations, typically one or two orders of magnitude shorter than a galaxy cluster. This allows us to test our implemented physics in a computationally efficient manner and, on the other hand, the large scale structure of the group/cluster is not expected to affect the micro-scale phenomena we have implemented as described in the previous section.

Following the analysis carried out in C26a, in this work we employ two different scenarios in terms of initial conditions at both strong (\high) and weak (\low) turbulence to better study SMBH feeding and feedback processes in realistic environments. For reference, the main parameters of the turbulence driving used in C26a are shown in Table~\ref{tab:setup}. 
Following B26a, we drive turbulence solenoidally at \(n_{\rm peak}=4\), corresponding to $l_{\rm inj}=L_{\rm box}/n_{\rm peak}$. This idealized forcing scale is motivated by the large-scale vortical stirring observed and inferred in cool-core hot halos, where AGN bubbles, jet-driven uplift, minor mergers, and sloshing generate subsonic motions over core scales. The purely solenoidal prescription allows us to isolate the role of turbulent mixing in CCA condensation and jet-regulated feeding \citep[][]{Fabian2012,McNamaraNulsen2012,Vazza2012,GaspariChurazov2013,Hofmann2016,Hitomi2016,Xrism2025}.

It is important to note that in this work we investigate a specific case of the above described injection mechanism. In particular, we limit our efforts to studying the effects and the consequences of a kinetically-dominated jet, i.e. setting the kinetic fraction to 1 in Eq.~\ref{eq:approx_tot_power}. 
This choice is motivated by the need to isolate how injected kinetic energy is redistributed into thermal, turbulent, and multiphase gas motions across resolved scales. In addition, we will make use of our findings in future companion papers in the {\sc BlackHoleWeather} series, in accordance with the project aim to link and study AGN accretion and feedback at different scales and phases as a whole, following a \textit{bottom-up}, first-principles approach \citep{Gaspari2020}. However, we also note that the current implementation does not account for magnetic fields and extra physics, such as relativistic effects and AGN winds, which could potentially alter the results. We defer the study of these limitations and caveats to a future work. 

\section{Results}\label{sec:results}

In this section, we present the main results of the analysis for the two reference simulations, \high and \low, introduced in \S\ref{sec:methods}. The purpose of this work is to investigate how AGN feedback, injected on the smallest resolved scales, propagates through and reshapes the surrounding medium under two different turbulence regimes. In particular, turbulence driving and the ensuing cascade regulate both the degree of mixing and the thermodynamic fluctuations of the gas \citep[e.g.][]{GaspariChurazov2013}. As a result, the \high run develops a stronger level of non-thermal support than the \low case, delaying the onset of local thermal instability and therefore postponing the formation of condensed structures. 
The production runs are initialized from pre-evolved turbulent atmospheres with the target velocity dispersions listed in Table~\ref{tab:setup}; radiative cooling and self-regulated jet feedback are then enabled at the restart time, which defines \(t=0\) Myr for the analysis. We define \train as the time, measured from this restart, at which the first cold clouds and filaments form, condense, and produce a sustained increase in sink accretion.

Throughout the analysis, the subgrid feedback prescription is kept fixed (see \S\ref{sec:methods}), so variations in jet power directly trace variations in \(\dot M_\bullet\). This allows us to focus on the resolved problem: how condensation, stirring, uplift, and meso-scale transport regulate the coupling between multiphase gas and SMBH feeding. Variations in the unresolved disc-to-horizon physics or jet geometry are deferred to future {\sc BlackHoleWeather} work.

Following the multiscale and multiphase framework adopted throughout this work, we divide the gas into four characteristic radial domains, defined primarily with respect to the meso-scale region where precipitation, jet-induced circulation, and inward transport are most directly coupled. As per {\sc BlackHoleWeather} framework \citep{Gaspari2020}, we refer to the \textit{micro} scale as the inner region where gas is already approaching the unresolved accretion zone ($r<0.1\ \mathrm{kpc}$), the \textit{meso} scale as the central transport and coupling layer between condensation and SMBH feeding ($0.1\ \mathrm{kpc}<r<1\ \mathrm{kpc}$), the inner \textit{macro} scale as the region where extended filaments, uplift, and halo circulation connect to the meso-scale ($1\ \mathrm{kpc}<r<10\ \mathrm{kpc}$), and the outer \textit{macro} scale as the large-scale hot-atmosphere reservoir ($r>10\ \mathrm{kpc}$). 
To track how the gas changes across these regions, we separate it into five temperature-defined phase proxies (tied to realistic observational bands): molecular-temperature gas ($T<2\times10^{2}\ \mathrm{K}$), cold atomic gas ($2\times10^{2}\ \mathrm{K}\leq T<1.6\times10^{4}\ \mathrm{K}$), warm gas ($1.6\times10^{4}\ \mathrm{K}\leq T<1.16\times10^{6}\ \mathrm{K}$), soft-X-ray-temperature gas ($1.16\times10^{6}\ \mathrm{K}\leq T<5.8\times10^{6}\ \mathrm{K}$), and hard-X-ray-temperature gas ($T>5.8\times10^{6}\ \mathrm{K}$).

As in C26a, we use the terms \emph{stormy}, \emph{rainy}, and \emph{cloudy} as compact descriptors of CCA weather states, again mainly in relation to how the meso-scale connects large-scale condensation to micro-scale feeding. A stormy state denotes spatially extended, strongly mixed, and bursty precipitation, in which condensation spreads from the meso-scale into the surrounding halo and drives intermittent feeding. A rainy state denotes a more compact and coherent condensation cycle, where the meso-scale remains efficiently connected to the central sink and sustains SMBH fueling. A cloudy state denotes a multiphase atmosphere in which cold gas is still present, especially on meso- and macro-scales, but its coupling to direct SMBH feeding is weakened. These labels summarize the combined thermodynamic, kinematic, and accretion behaviour; they are not separate simulation inputs.

The results are presented as follows. We first study the time and space variability and accretion properties in \S\ref{subsec:variability}. In \S\ref{subsec:inflow_outflow} we investigate the inflow and outflow properties of the multiphase gas. The kinematic state of gas is described in \S\ref{subsec:kplot} under different BH weather conditions. The \cratio-ratio profiles are finally discussed in \S\ref{subsec:c-ratio} to probe different CCA regimes.

\subsection{Variability \& accretion properties}\label{subsec:variability}

SMBH accretion is intrinsically time-variable, and in a self-regulated feedback model this variability directly modulates the jet power. Observationally, such variability can appear in optical/UV continuum luminosity changes and in the broadening or narrowing of emission-line profiles \citep{TozziLusso2022,CammelliTan2025,PaolilloPapadakis2025}. In the CCA framework, this flickering behaviour is expected to arise from the stochastic arrival, interaction, and partial angular-momentum cancellation of multiphase clouds and filaments near the unresolved accretion region.
In our simulations, gas accreted through the pc-scale sink is mapped onto the unresolved central engine through the prescription described in \S\ref{sec:methods}. We therefore use \(\dot M_\bullet(t)\) as a tracer of the time-dependent CCA feeding cycle and of the corresponding jet-power response. As shown in the no-jet simulations of B26b, CCA feeding naturally produces scale-dependent variability: the relative amplitude of \(\dot M_\bullet\) fluctuations increases with the temporal window over which the accretion history is sampled, producing flicker-like accretion histories rather than white-noise fluctuations. Here we examine how this intrinsic CCA variability is modified once the accretion rate is coupled to a self-regulated kinetic jet in the \texttt{high-turb} and \texttt{low-turb} reference runs.

\subsubsection{Feeding the SMBH}
\begin{figure}
    \centering
    \includegraphics[width=1\linewidth]{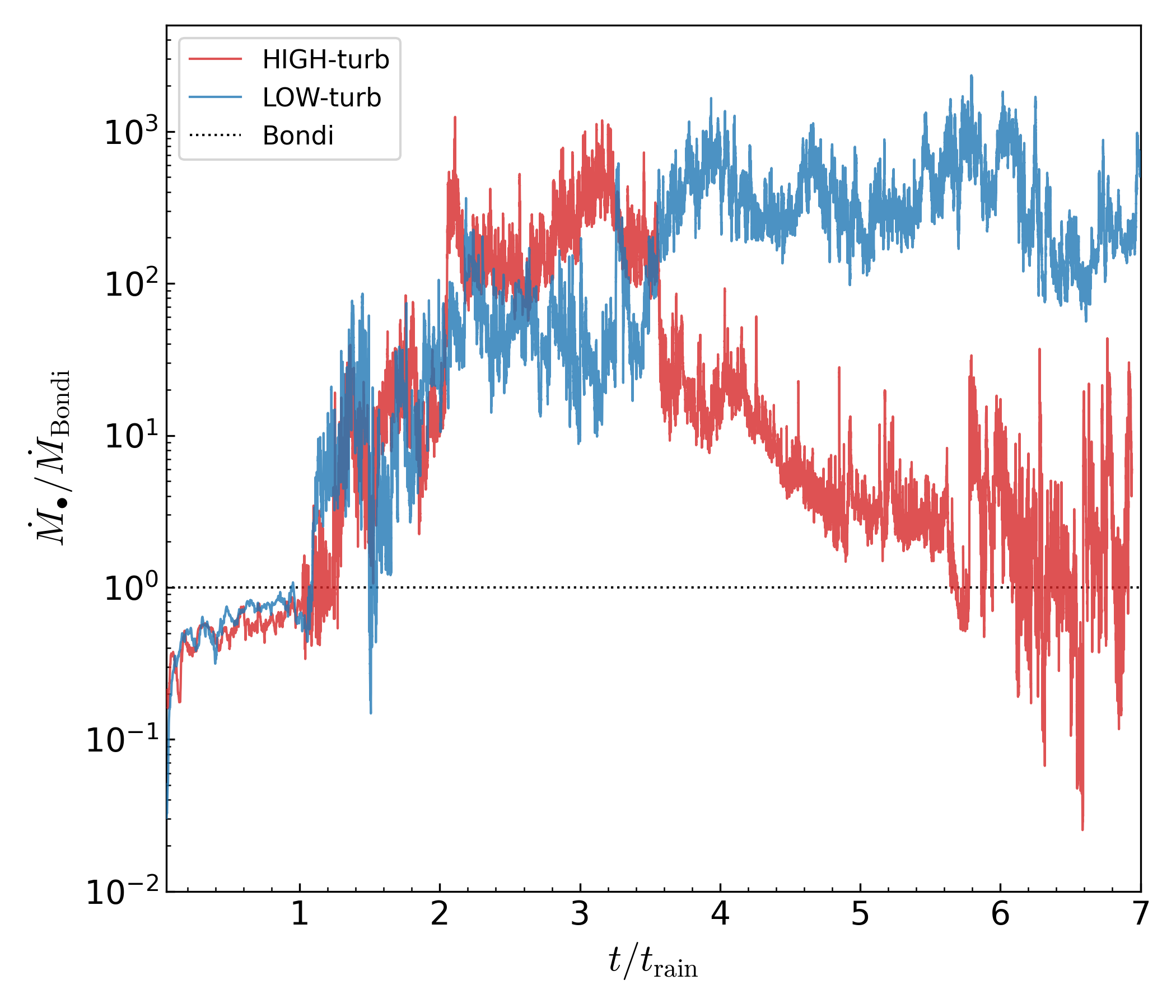}
    \caption{Black hole accretion rate as a function of time in the \low and \high case, as indicated in the legend. The simulation time has been normalized to visually compare the different runs starting from the time they have been restarted with the jet and the cooling on. During active jet phases, \bhar reaches large super-Bondi values (cf.~B26a,b). The dotted horizontal line marks the Bondi reference accretion rate, $\dot M_\bullet/\dot M_{\rm Bondi}=1$.}
    \label{fig:bhar}
\end{figure}

Figure~\ref{fig:bhar} shows the accretion-rate evolution of the two reference runs, \high\ and \low, in units of the corresponding Bondi accretion rate, \(\dot{M}_{\rm Bondi}\), with time normalized to \(t_{\rm rain}\). This normalization provides a useful hot-mode baseline \citep{Bondi1952}: values close to unity indicate accretion rates comparable to smooth, quasi-spherical capture from the hot atmosphere, whereas strongly super-Bondi values mark the transition to multiphase, precipitation-driven feeding. Bondi-like prescriptions have been widely adopted as subgrid SMBH accretion models in cosmological and galaxy-scale simulations \citep[e.g.][]{SpringelDiMatteoHernquist2005,BoothSchaye2009}, and have also been used as observationally motivated estimates of hot-mode feeding in nearby systems \citep[e.g.][]{AllenDunn2006}. 

In the present CCA context, however, \(\dot{M}_{\rm Bondi}\) should be interpreted only as a reference scale rather than as a predictive accretion model, because the flow is turbulent, multiphase, anisotropic, and directly affected by jet feedback. We define
\begin{equation}
\dot{M}_{\rm Bondi}
= 4\pi\,\lambda(\gamma)\,\frac{(G M_\bullet)^2\,\rho_\infty}{c_{s,\infty}^{\,3}} \, ,
\end{equation}

where \(\lambda(\gamma)\) is a dimensionless factor of order unity that depends on the adiabatic index $\gamma$ (here assumed 5/3), while \(\rho_\infty\) and \(c_{s,\infty}\) are the ambient density and sound speed measured in the hot atmosphere at large radii. The assumptions entering this expression, namely a smooth, homogeneous, radially symmetric, and adiabatic hot inflow, are not satisfied in realistic jet-regulated CCA atmospheres. For this reason, and consistently with B26a,b, we use \(\dot{M}_{\rm Bondi}\) only to quantify departures from the idealized hot-mode limit and to identify when precipitation boosts the accretion rate into a genuinely multiphase feeding regime.

During the first normalized rain time, both runs evolve in a similar way, slowly rising from sub-Bondi values toward $\dot{M}_\bullet/\dot{M}_{\rm Bondi}\sim1$. The \high\ case remains slightly lower than \low\ in this initial stage, consistent with the stronger large-scale stirring already discussed in C26a, which enhances mixing and temporarily hinders the delivery of gas to the sink.  At these early times, the jet has not yet had enough time to reshape the inner multiphase structure in a substantial way. Once condensation sets in, both runs move rapidly into a super-Bondi regime, marking the transition from smooth hot feeding to genuine CCA. The subsequent evolution closely mirrors the thermodynamic ``weather'' states identified in C26a. In the \high\ run, the accretion ratio rises abruptly after $t/t_{\rm rain}\sim1$ and reaches a short but intense super-Bondi episode, peaking at $\dot{M}_\bullet/\dot{M}_{\rm Bondi}\sim10^{2}$--$10^{3}$ around $t/t_{\rm rain}\sim2$--3.5. This is the accretion counterpart of the early \stormy phase: the atmosphere hosts vigorous precipitation, broad phase mixing, and rapid delivery of dense condensates toward the nucleus. After $t/t_{\rm rain}\sim3.5$, however, the accretion rate declines by more than an order of magnitude and becomes increasingly intermittent, reaching on average values slightly larger or comparable to the Bondi case. This behaviour is consistent with the later \cloudy phase identified in the previous paper, in which cold clouds and filaments are still present at micro- and meso-scales but are less efficiently channeled into the sink region.  In other words, the high-turbulence atmosphere continues to form multiphase structure, but its late-time precipitation is less effective at sustaining direct SMBH feeding.

The \low\ run follows a different path.  After the initial transition to super-Bondi accretion, the feeding history remains elevated for a much longer interval, with $\dot{M}_\bullet/\dot{M}_{\rm Bondi}$ persistently in the $\sim10^{2}$--$10^{3}$ range over several $t_{\rm rain}$.  This prolonged active state is fully consistent with the results of C26a, which showed that the low-turbulence run retains a denser and more persistent molecular/cold reservoir within the inner kpc.  In this calmer atmospheric state, condensation is less explosive but more effectively retained near the centre, so that the SMBH continues to be fueled over an extended period rather than through a single dominant burst.  Thus, while the \high\ run produces an earlier and more impulsive accretion episode, the \low\ run supports a longer-lived, reservoir-fed mode of CCA.  Read together with the multiphase diagnostics of C26a, Figure~\ref{fig:bhar} shows that ambient turbulence controls not only the onset of condensation, but also how efficiently the resulting cold gas can feed the SMBH: either through short-lived \textit{storm}-fed accretion episodes or through a longer-lasting \rainy, reservoir-supported mode.
Thus, the key distinction is not whether condensation occurs in both runs, but whether the resulting cold gas remains dynamically connected to the central sink.

\subsubsection{Accretion-rate spectral analysis}\label{subsec:spectrum}
\begin{figure}
    \centering
    \includegraphics[height=0.823\textheight, width=1\linewidth]{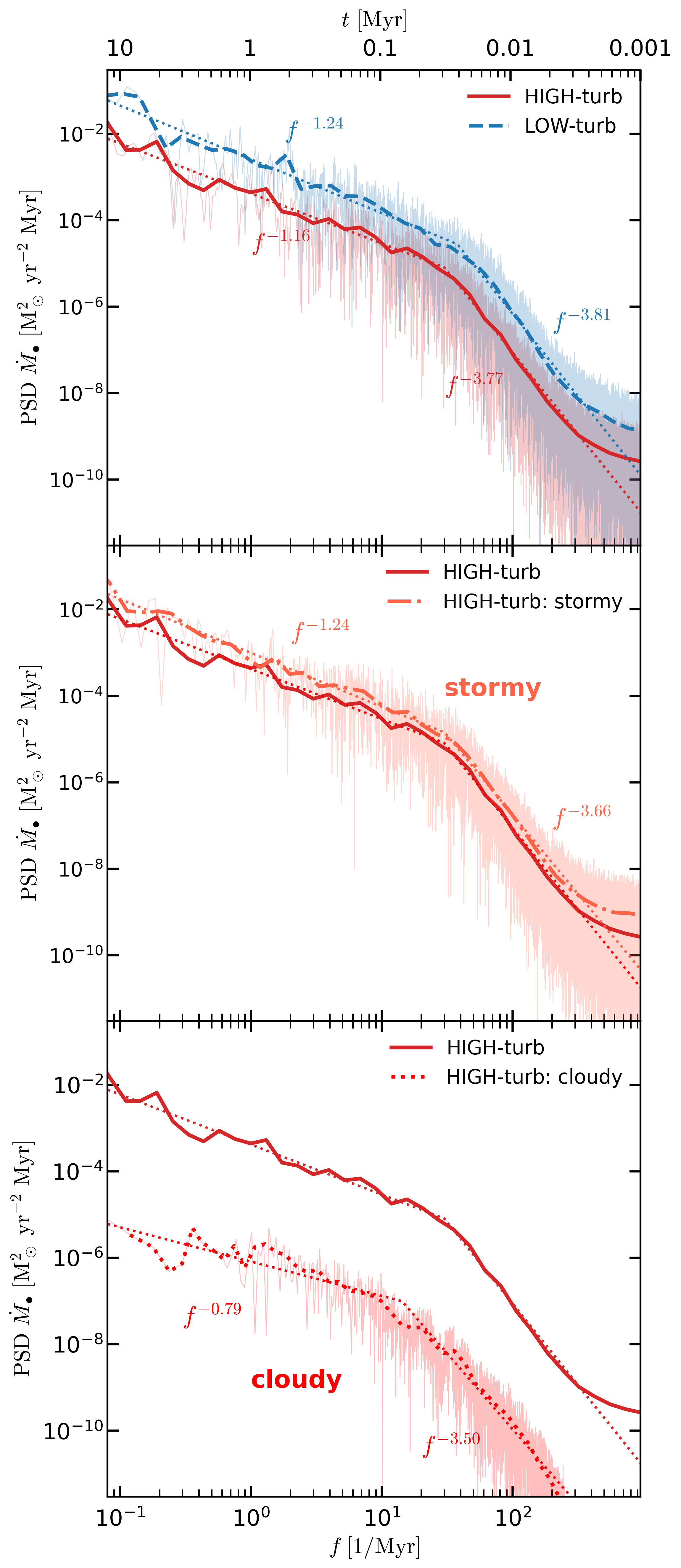}
    \caption{Power spectral density (PSD) as a function of frequency in the \low and \high case, as indicated in the legend, from the time the jet and the cooling have been restarted. Light solid lines show the direct PSD from the \bhar  reported in Figure~\ref{fig:bhar}, while dark solid lines depict the smoothed version using a Gaussian filter. Dashed lines instead represent a broken power law fit to the dark solid lines. Coloured text indicates the correspondent slope of the fitted curves in separate regimes. Upper panel: full ($t/t_{\rm rain} = 1-7$) evolution of the \high and \low cases, with a low-frequency slope close to $f^{-1.2}$ and a much steeper high-frequency tail, $\sim f^{-3.8}$. Middle/bottom panels: the full \high PSD is compared to the \stormy and \cloudy regime, respectively.}
    \label{fig:psd}
\end{figure}

\begin{table}
\caption{Broken power-law fits to the accretion-rate PSD. We fit the smoothed PSD with
\(
P(f)=A
\begin{cases}
\left(f/f_{\rm b}\right)^{\alpha_{\rm 1}}, & f < f_{\rm b} \\
\left(f/f_{\rm b}\right)^{\alpha_{\rm 2}}, & f \ge f_{\rm b}
\end{cases}
\),
where $f_{\rm b}$ is the break (knee) frequency. Reported uncertainties correspond to the formal $1\sigma$ fit errors.}
\label{tab:psd}
\centering
\setlength{\tabcolsep}{3pt}
\begin{tabular}{lccc}
\toprule
Run & $\alpha_1$ & $\alpha_2$ & $\log_{10}(f_b/[{\rm Myr}^{-1}])$ \\
\midrule
\low & $-1.24 \pm 0.06$ & $-3.81 \pm 0.19$ & $1.55 \pm 0.06$ \\
\high & $-1.16 \pm 0.05$ & $-3.77 \pm 0.15$ & $1.48 \pm 0.05$ \\
\texttt{high-}\stormy & $-1.24 \pm 0.04$ & $-3.66 \pm 0.12$ & $1.48 \pm 0.04$ \\
\texttt{high-}\cloudy & $-0.79 \pm 0.09$ & $-3.50 \pm 0.20$ & $1.15 \pm 0.07$ \\
\hline
\end{tabular}
\end{table}

To characterize the variability of the accretion flow beyond the time-domain behaviour, we compute the power spectral density (PSD) of $\dot{M}_\bullet$, shown in Figure~\ref{fig:psd}. The PSD is calculated, in a given simulation time range, as the square modulus of the Fourier transform per frequency unit of the \bhar. We restrict the displayed and fitted range to \(f\lesssim10^3~{\rm Myr}^{-1}\).\footnote{Although the \bhar is output at every integration time step, $\sim3\times10^{-6}$ Myr, the PSD at higher frequencies is dominated by a quasi white-noise flattening arising from the finite numerical resolution of the simulation.} Solid thin transparent lines indicate the pure PSD of the \bhar in the frequency space, while thick lines denote the average interpolated binned PSD in the corresponding spectral range. We attempt to fit the interpolated PSD by using a double power-law function, which parameters and relative errors are reported in Table~\ref{tab:psd}. The upper panel compares the full \bhar time series of the \high and \low runs over the range 1\,-\,$7\ t_{\rm rain}$, once radiative cooling sets in. In both cases the PSD is well described by a broken power law, with a low-frequency slope close to $f^{-1.2}$ and a much steeper high-frequency tail, $f^{-3.8}$. This implies that the accretion variability is not white-noise dominated, but instead displays a flicker-like/pink-noise behaviour (cf.~\citealt{Gaspari2017}) on long timescales, which steepens into a strongly red spectrum at high frequencies, as also found in the pure CCA feeding simulations (B26b). The two runs therefore share a similar underlying variability cascade, but differ substantially in normalization: the \low run carries systematically more power at nearly all frequencies, consistent with its more persistent super-Bondi feeding history and longer-lived central cold reservoir. The \high run, instead, shows a lower overall PSD normalization once the full evolution is considered, reflecting the fact that its early bursty accretion phase is followed by a later suppression of direct SMBH feeding. 

We also note that the two cases show remarkably similar break frequencies, $f_{\rm b}\simeq30\ \rm Myr^{-1}$ ($\rm log (f_{\rm b}/[{\rm Myr}^{-1}])\simeq 1.48$) in the \high and $f_{\rm b}\simeq37.8\ \rm Myr^{-1}$ ($\rm log (f_{\rm b}/[{\rm Myr}^{-1}])\simeq 1.55$) in the \low, corresponding to break times of $t_{\rm b}\simeq\ 0.033\ \rm Myr$ and $t_{\rm b}\simeq\ 0.026\ \rm Myr$, respectively. This indicates that the typical timescales for the multiphase gas to lose temporal coherence (likely due to cloud collisions) result similarly in the two reference cases. This behaviour differs, in terms of break frequency, from the study presented in B26b where $f_{\rm b}$ shows a remarkable separation due to the strength of the turbulence driving ($\simeq9\ \rm Myr^{-1}$ in the low-turbulence scenarios versus $\simeq53\ \rm Myr^{-1}$ of the high-turbulence case). This suggests that in the case of jet-regulated CCA, the dependence on the turbulence driving strength is somehow alleviated by the extra term introduced by the jet turbulence itself, which affects the collisional dynamics of multiphase clouds and filaments, and therefore impacts their accretion signal down to the central SMBH. As discussed in B26b, the implications of $t_{\rm b}$ are physically motivated when considering that $t_{\rm b}$ in our simulations is much longer than the integration timestep over which \bhar is updated ($\sim 10^{-6}\ \rm Myr$), and far shorter than the correlation time implemented in the turbulence driving (see Table~\ref{tab:setup},  $t_{\rm corr} = 30\ \rm Myr$).

This interpretation becomes clearer when the two weather states identified in the \high run are isolated. In the middle panel, the \stormy interval retains a low-frequency slope of $\propto f^{-1.24}$ , i.e. close to pink noise (cf.~\citealt{Gaspari2017}), together with a steeper red-noise high-frequency tail ($\propto f^{-3.66}$) and a consistent $f_{\rm b}\simeq30\ \rm Myr^{-1}$. This indicates a regime of correlated, scale-coupled accretion variability, driven by cloud-cloud collision, in which precipitation events remain linked across a broad range of timescales. In the lower panel, the \cloudy interval is instead characterized by a much lower overall normalization and a flatter low-frequency slope, $\propto f^{-0.79}$, i.e. intermediate between white and pink noise, while the high-frequency tail retains the trend of a red noise ($\propto f^{-3.50}$). A possible interpretation is that the \cloudy phase affects the long-timescale accretion variability: the large, coherent rain episodes typical of the \stormy state are replaced by weaker and less correlated feeding events, even though rapid small-scale flickering is still present. This behaviour is consistent with the thermodynamic picture from C26a, in which the \stormy phase corresponds to vigorous precipitation directly feeding the nucleus, whereas the \cloudy phase still forms cold clouds and filaments on micro- and meso-scales but couples them less efficiently to the sink.

A complementary comparison is provided by P26b, where jet feedback includes spin-dependent precession and the turbulence-driving history is varied. In those simulations, the low-frequency slopes also remain close to \(P(f)\propto f^{-1}\), while the main differences appear in the PSD normalization and in the high-frequency damping. This suggests that the low-frequency flicker-noise regime is common to the different CCA realizations, whereas the detailed PSD shape depends on the specific coupling between turbulence, jet feedback, and the multiphase inflow.


\subsubsection{Eddington ratios}

\begin{figure}[!ht]
    \centering
    \includegraphics[width=1\linewidth]{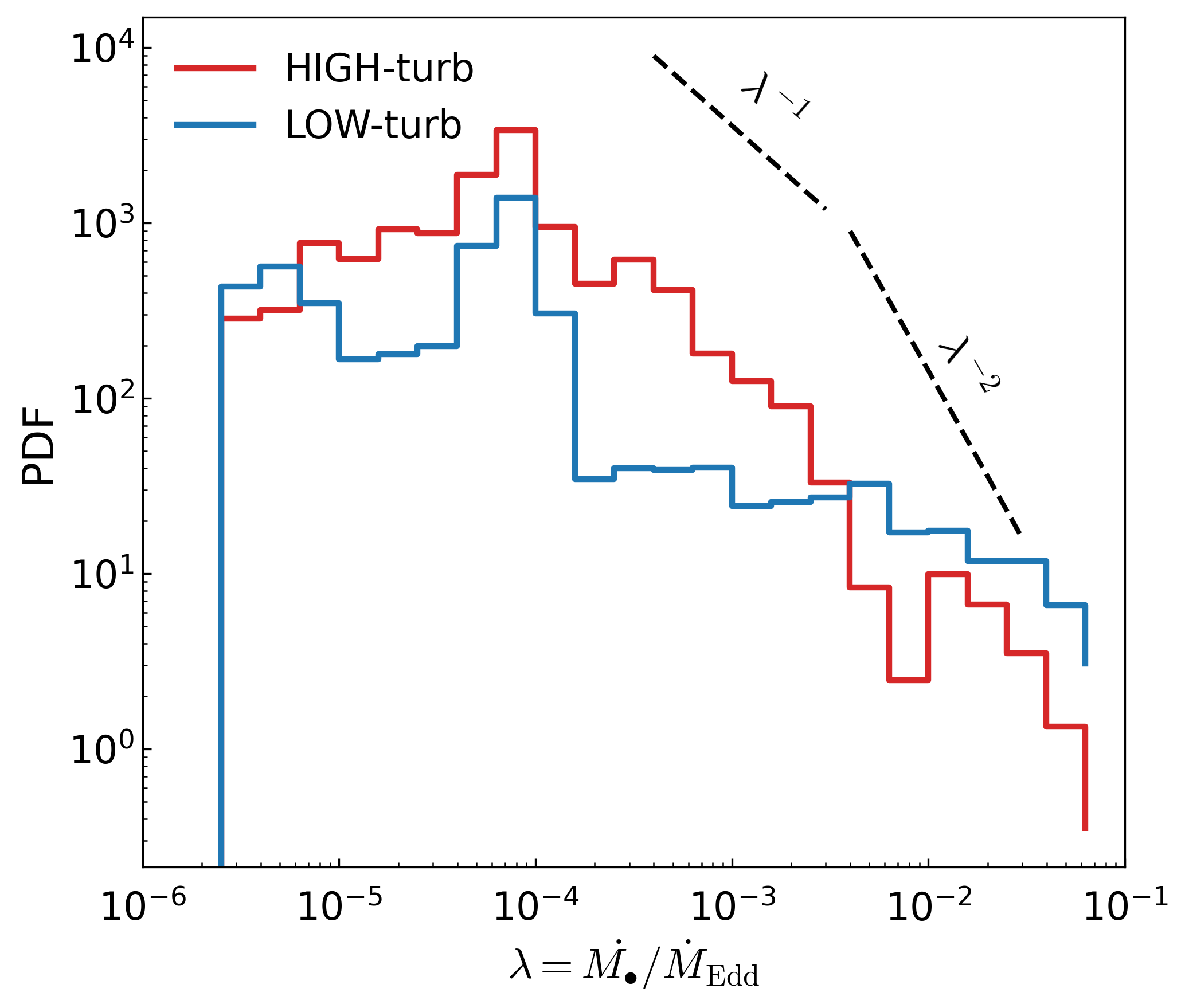}
    \caption{Probability density function (PDF), normalized to unity in logarithmic space, of the Eddington ratios $\lambda_{\rm Edd} \equiv \dot{M_{\bullet}}/\dot{M}_{\rm Edd}$ as derived from the \bhar for the 2 reference runs, \high (red solid line) and \low (blue solid line), computed at each time step. Black dashed guide lines indicate approximate slopes of $-1$ and $-2$ in the high-$\lambda_{\rm Edd}$ tail. }
    \label{fig:lambda}
\end{figure}

To fully characterize the accretion properties of our simulations, it is useful to study not only the absolute contribution of the \bhar but also its role relative to the mass of the central BH. Figure~\ref{fig:lambda} shows the probability density function (PDF) of the BH accretion rates normalized to the Eddington accretion rate $\dot{M}_{\rm Edd}$, corresponding to the Eddington luminosity $L_{\rm Edd} = 4\pi G M_{\bullet} m_{\rm p} c / \sigma_T$, where $\sigma_T = 6.65\times10^{-25}\,{\rm cm}^2$ is the Thomson scattering cross section, $m_{\rm p} = 1.67\times10^{-24}\,{\rm g}$ is the proton mass and $c = 3\times10^{10}\,{\rm cm\,s^{-1}}$ is the speed of light. Note that the PDF is obtained using equally spaced logarithmic bins, such that the integral in  $d\ log(\lambda)$ is equal to 1. With a fiducial SMBH mass \(M_\bullet=2.8\times10^8\,M_\odot\) and a reference radiative efficiency \(\epsilon_{\rm rad}=0.1\), this gives
\begin{equation}
\dot M_{\rm Edd}=\frac{L_{\rm Edd}}{\epsilon_{\rm rad} c^2}\simeq6.2\ M_\odot\,{\rm yr}^{-1}.    
\end{equation}

The ratio $\lambda_{\rm Edd} \equiv \dot{M_{\bullet}}/\dot{M}_{\rm Edd}$ provides a dimensionless metric that is key to evaluating the role of the central engine in relative terms, and relatively easy to access in different spectral bands.
Observationally, \(\lambda_{\rm Edd}\) is usually inferred as \(L_{\rm bol}/L_{\rm Edd}\), rather than as a direct mass-accretion-rate ratio. The bolometric luminosity is commonly estimated either from multi-wavelength SED modelling or from band-dependent bolometric corrections applied to optical/UV, X-ray, infrared, or emission-line tracers \citep[e.g.][]{Lamastra2009,Duras2020, GuptaRicci2025}, while \(M_\bullet\) is inferred from reverberation-mapped or single-epoch virial estimators in unobscured AGN, and from host-galaxy scaling relations, dynamical modeling, or indirect X-ray/line proxies in obscured and low-luminosity systems \citep[e.g.][]{VestergaardPeterson2006}. Our simulated \(\lambda_{\rm Edd}\) should therefore be interpreted as an accretion-rate analogue rather than as a forward-modeled luminosity ratio, since converting \(\dot M_\bullet\) into \(L_{\rm bol}\) would require assumptions about radiative efficiency, accretion mode, obscuration, and bolometric corrections.

Focusing on our fiducial \high\ and \low\ runs, both distributions peak at \(\lambda_{\rm Edd}\simeq8\times10^{-5}\), extend down to \(\lambda_{\rm Edd}\sim10^{-6}\), and develop high-\(\lambda_{\rm Edd}\) tails reaching rare episodes up to \(\lambda_{\rm Edd}\sim10^{-1}\), as expected in mechanically-dominated AGN feedback \citep[e.g.][]{Russell2013}. Overall, the simulated feeding histories span nearly five orders of magnitude, consistent with the broad, strongly time-variable accretion-rate distributions inferred for AGN populations \citep[e.g.][]{KauffmannHeckman2009,Aird2012}. The low absolute normalization mainly reflects the large SMBH mass adopted in our setup, \(M_\bullet=2.8\times10^{8}\,M_{\odot}\) (Table~\ref{tab:setup}), so that even substantial sink accretion corresponds to modest Eddington-scaled rates. In addition, the relatively short simulated time span, \(\lesssim100~{\rm Myr}\), limits the sampling of very rare near-Eddington events that would be better captured over Gyr-scale duty cycles. Both runs therefore remain predominantly in a low-Eddington, mechanically dominated regime, broadly consistent with radio-mode AGN activity \citep[e.g.][]{BestHeckman2012,HeckmanBest2014}, while still showing intermittent excursions toward higher accretion states relevant to wind/feedback-regulated systems. 

\begin{figure}
    \centering
    \includegraphics[width=1\linewidth]{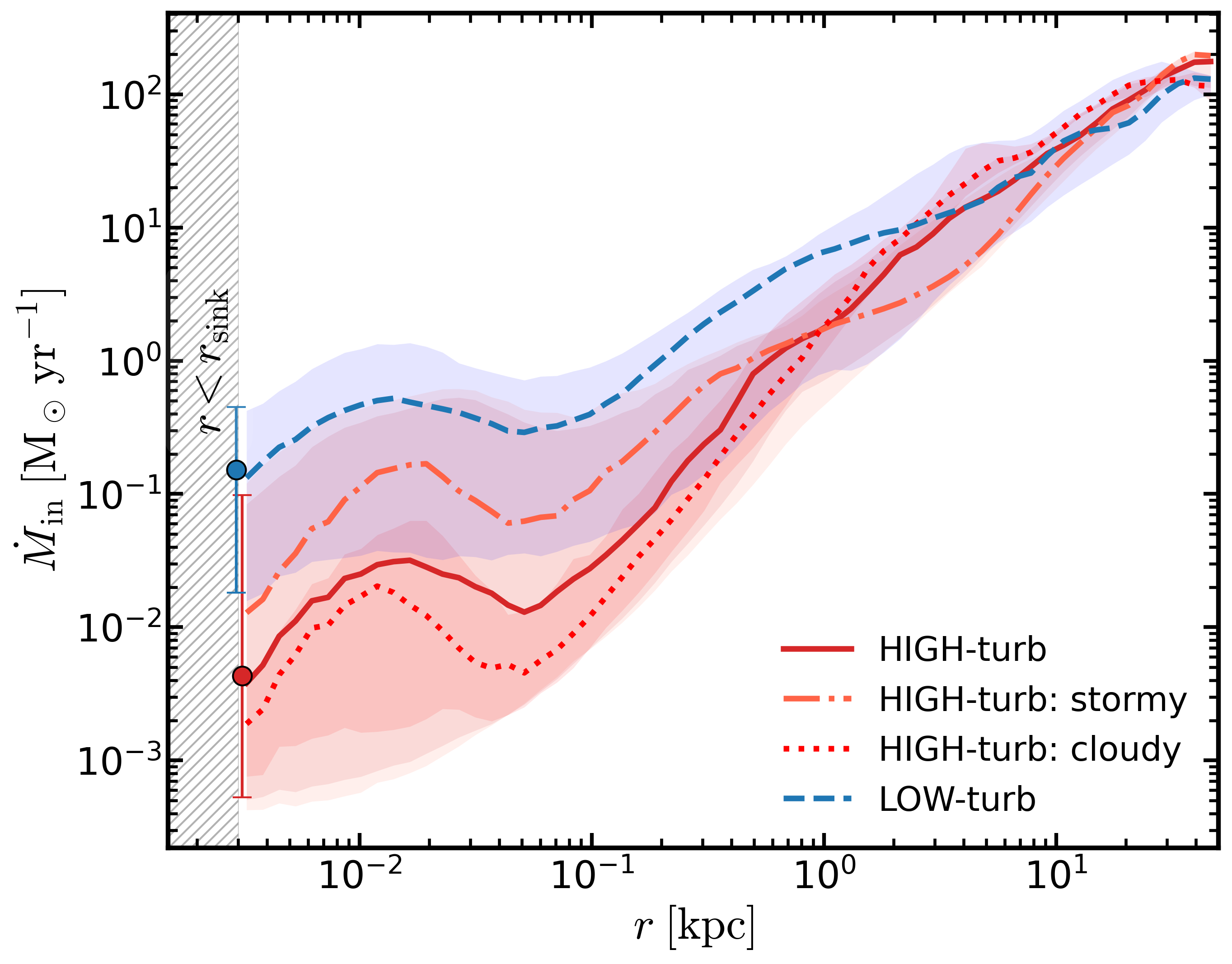}
    \caption{Inflow rate as a function of the radius in the \low and \high case, as indicated in the legend. For the latter, we subdivide the two different weather phases, \stormy (dash-dotted) and \cloudy (dotted lines). Lines show the 50th percentile throughout the whole duration of the simulation, with the shaded areas indicating the 1 sigma levels at any given radius (16th and 84th percentiles). For comparison, colored circles at $r=r_{\rm sink}$ display the time median of \bhar for the \low (blue) and \high (red) cases.}
    \label{fig:accr_radial}
\end{figure}

Despite their similar peak values, the two reference runs differ in the shape of the high-\(\lambda_{\rm Edd}\) tail. In the \high\ run, the distribution is approximately skewed toward large \(\lambda_{\rm Edd}\), with a logarithmic slope close to \(-1\) over \(\lambda_{\rm Edd}\sim10^{-4}-10^{-2}\), reminiscent of the scale-free, flicker-like accretion/feedback variability discussed in micro-scale disk-wind models \citep{Fiore2024}, before steepening to \(\sim-2\) for \(\lambda_{\rm Edd}\gtrsim10^{-2}\). By contrast, the \low\ run displays a flatter tail, with slopes shallower than \(-1\) across the sampled \(\lambda_{\rm Edd}\) range. This indicates that, although both systems remain in a low-Eddington state on average, the \low\ atmosphere more frequently sustains relatively high Eddington-scaled feeding episodes, consistent with its larger \(\dot M_\bullet/\dot M_{\rm Bondi}\) and stronger inner inflow (see Figures~\ref{fig:bhar} and \ref{fig:accr_radial}, in the next section). The flatter PDF in \low, which is not seen in the no-jet B26b comparison, suggests that jet-regulated CCA modifies the mapping between turbulence and SMBH feeding. As also suggested by the morphological maps in C26a, weaker ambient stirring plausibly reduces cloud and filament fragmentation, limiting meso-scale decoupling and allowing condensates across a broader mass range to remain connected to the central accretion region.

\subsection{Inflow \& Outflow}\label{subsec:inflow_outflow}

\begin{figure}
    \centering
    \includegraphics[width=1\linewidth]{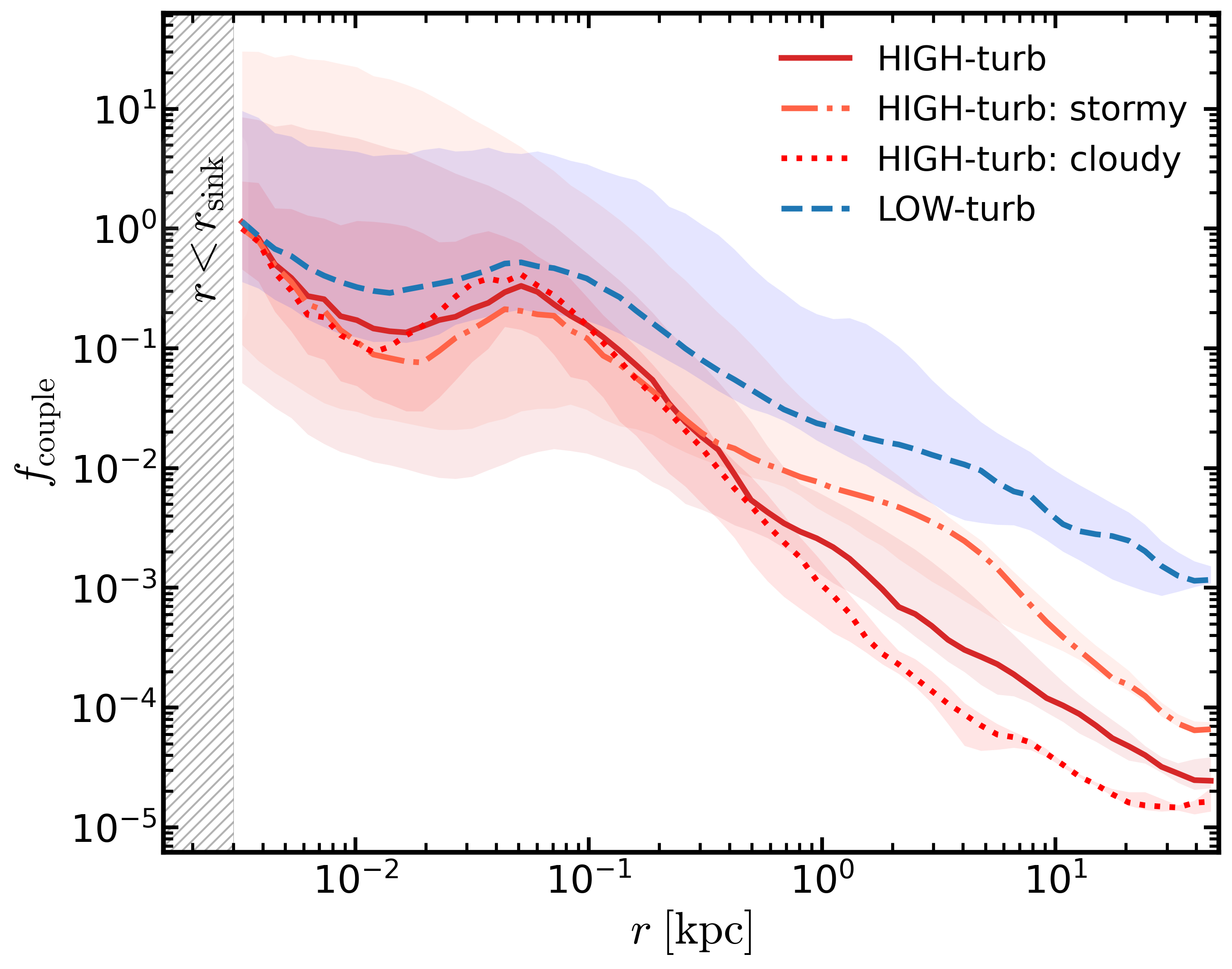}
    \caption{Inflow coupling efficiency as a function of the radius in the \low (blue dashed line) and \high (red solid line) case. As in Figure~\ref{fig:accr_radial}, lines indicate the 50th percentile throughout the total duration of the simulation, with the shaded areas showing, at any given radius, the 1 sigma levels (16th and 84th percentiles).}
    \label{fig:f_couple}
\end{figure}

\begin{figure*}
    \centering
    \includegraphics[width=0.75\linewidth]{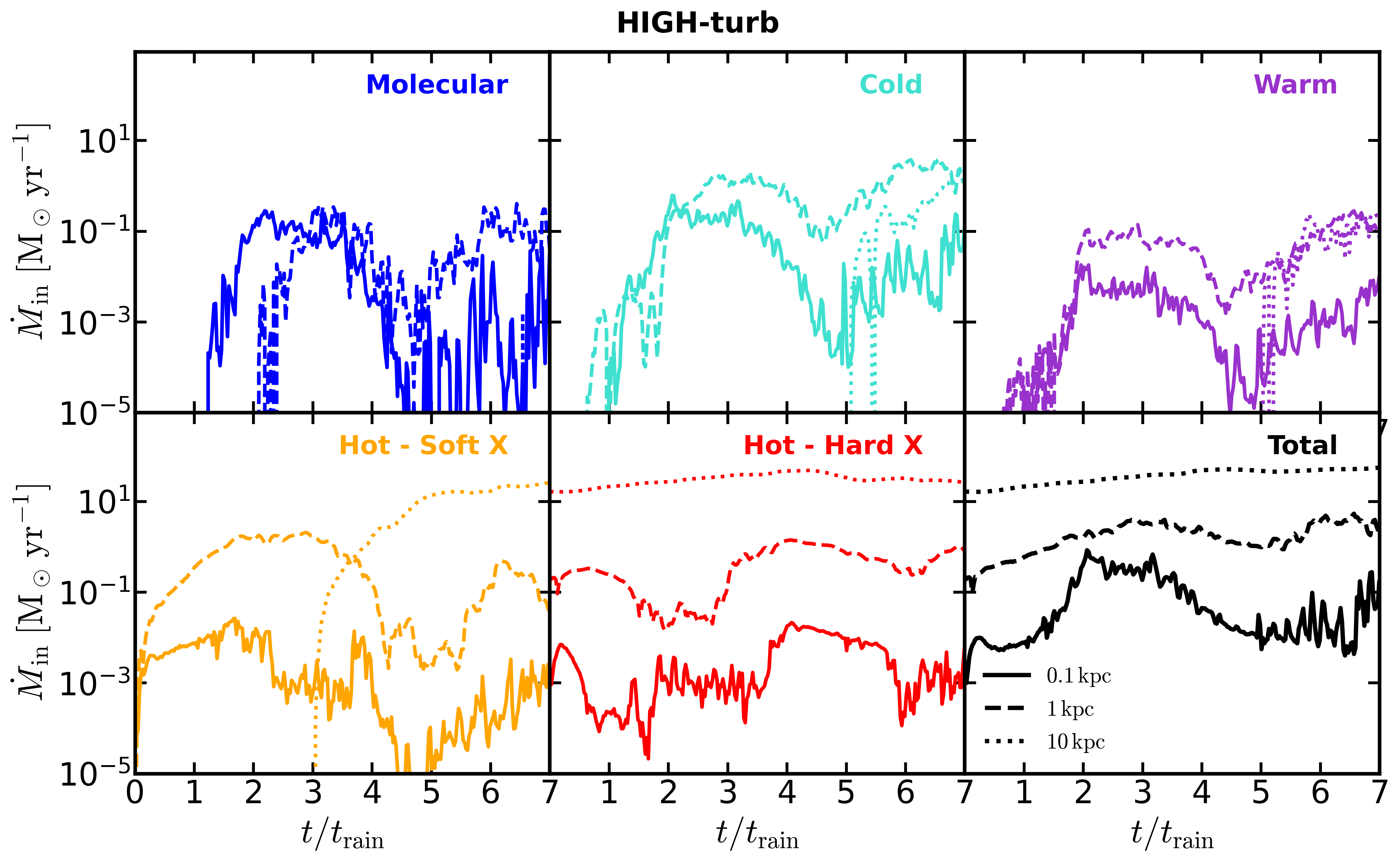}
    \includegraphics[width=0.75\linewidth]{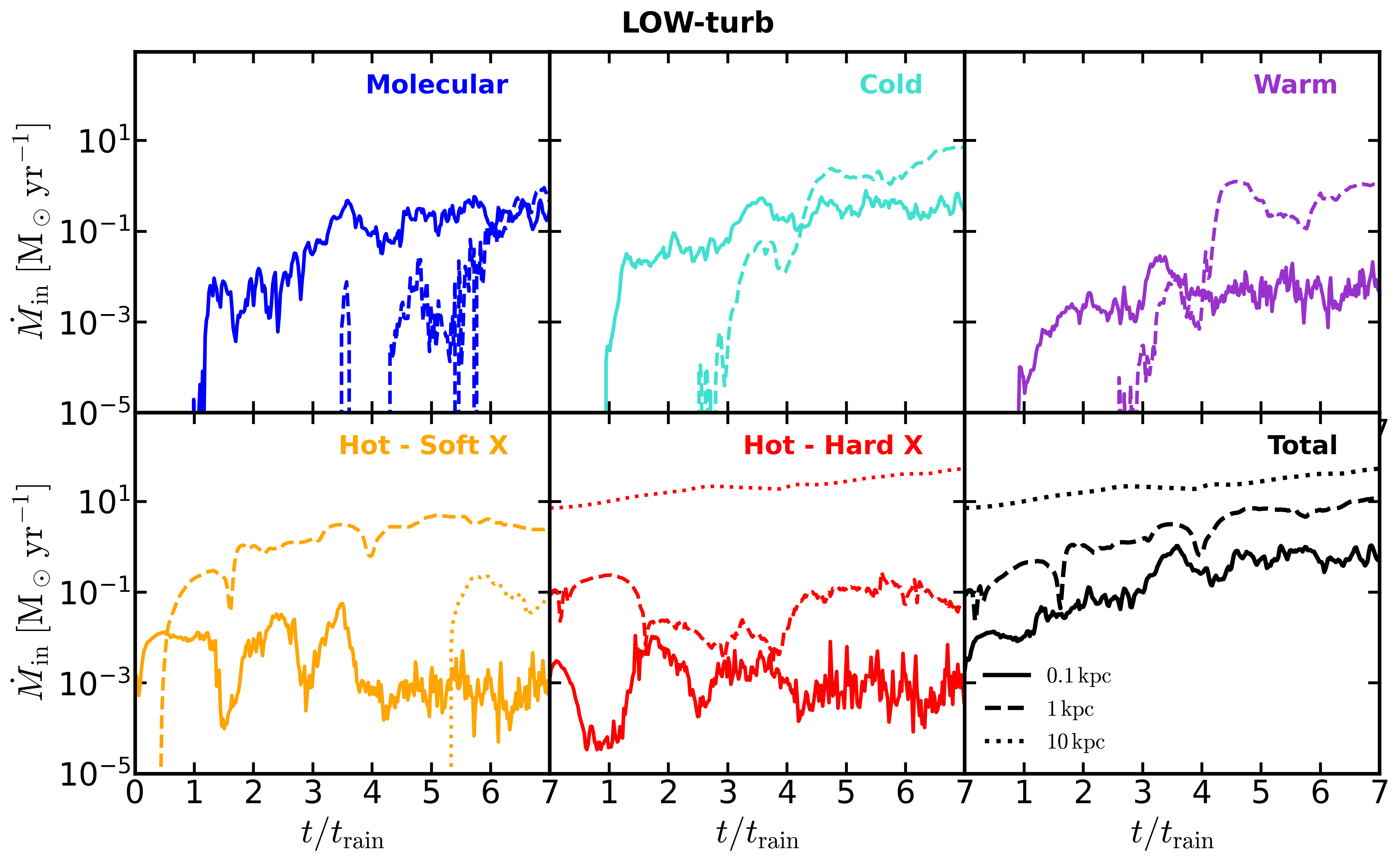}
    \caption{Phase-separated mass inflow rates for the \high (top figure) and \low (bottom figure) runs as a function of time normalized to $t/t_{\rm rain}$.  Inflow is defined from gas with negative radial velocity, and each panel shows the corresponding $\dot{M}_{\rm in}$ for a given thermal phase: molecular, cold, warm, soft X-ray, hard X-ray, and the total (black lines).  Solid, dashed, and dotted lines refer to $r=0.1$, 1, and 10~kpc, respectively. In the \high scenario, once condensation is established, the inflow becomes increasingly multiphase on micro- and meso-scales, while the outer halo remains dominated by the hot phases. The \low condensed inflow remains more persistent at late times, consistent with a longer-lived central cold reservoir and sustained SMBH feeding.}
    \label{fig:inflow}
\end{figure*}

\begin{figure*}
    \centering
    \includegraphics[width=0.75\linewidth]{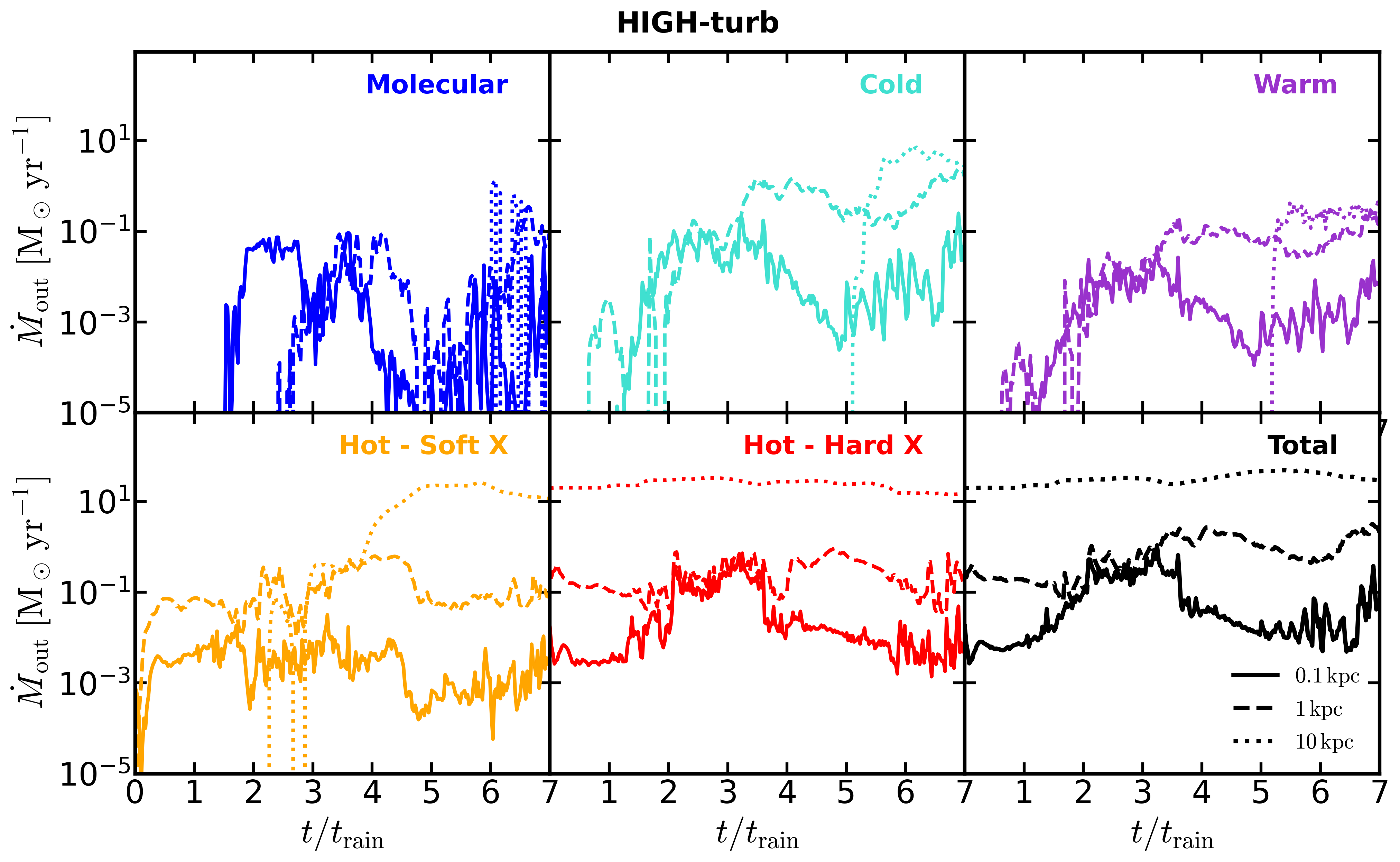}
    \includegraphics[width=0.75\linewidth]{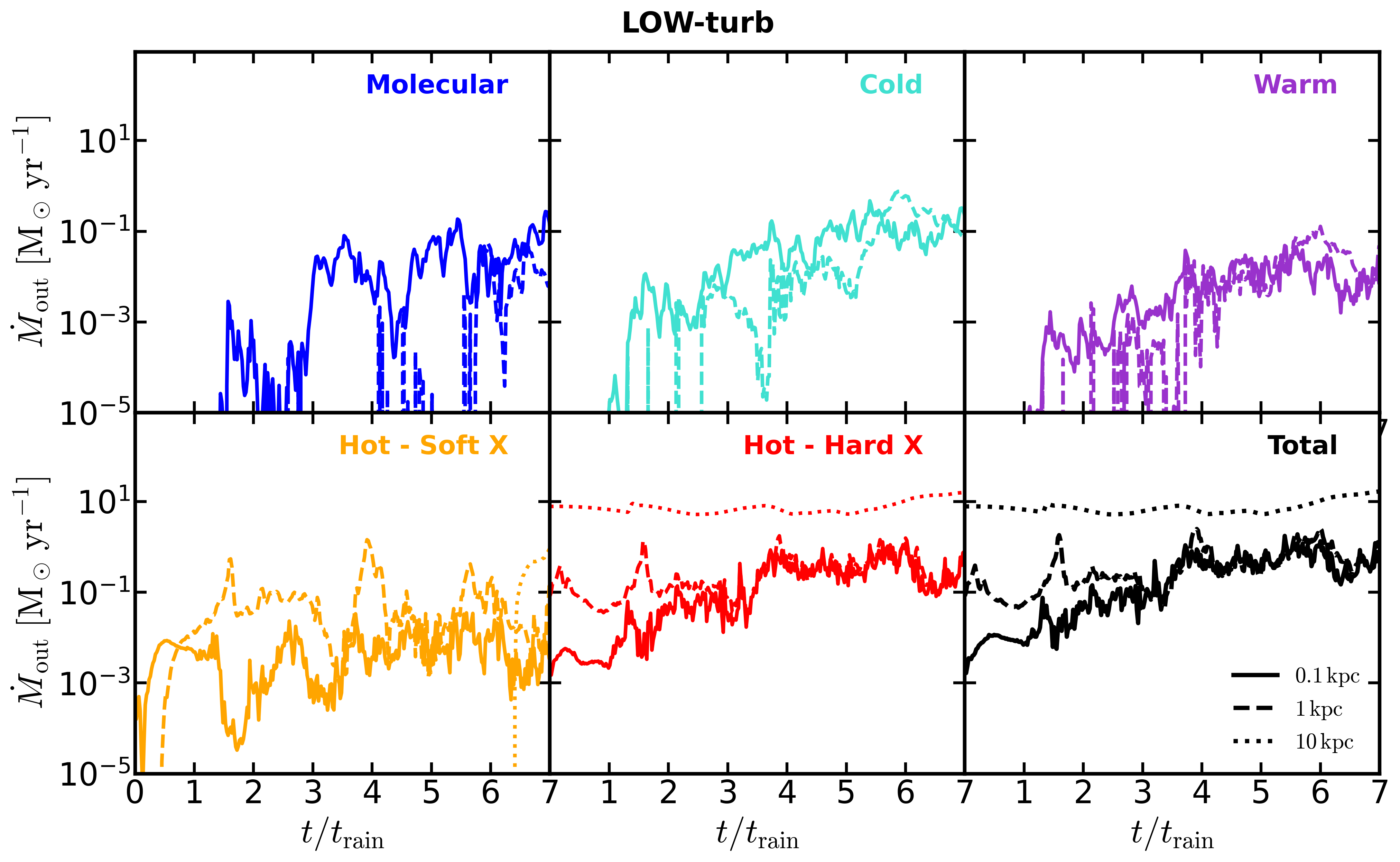}
    \caption{Phase-separated mass outflow rates for the \high (top figure) and \low (bottom figure) runs as a function of time normalized to $t/t_{\rm rain}$.  Outflow is defined from gas with positive radial velocity, and each panel shows the corresponding $\dot{M}_{\rm out}$ for a given thermal phase: molecular, cold, warm, soft X-ray, hard X-ray, and the total (black lines). Solid, dashed, and dotted lines refer to $r=0.1$, 1, and 10~kpc, respectively.  In the \high case, the molecular, cold and warm outflows are more extended and intermittent, consistent with stronger uplift and entrainment by the more disturbed medium. In the \low atmosphere, cool outflows are mostly confined to the inner kpc and are less intermittent, indicating a lower entrainment/uplifting due to the weaker turbulent stirring efficiency.}
    \label{fig:outflow}
\end{figure*}

In order to further characterize the CCA cycle, in this section we study the properties of the inflowing and outflowing material as a function of time and radius. Operatively, we classify as \emph{inflow}/\emph{outflow} each gas cell that has negative/positive radial velocity with respect to the SMBH, respectively. At radius \(r\), we define:
\[
\dot M_{\rm in}(r)=\int \rho\,\max(-v_r,0)\,dA,
\]
and
\[
\dot M_{\rm out}(r)=\int \rho\,\max(v_r,0)\,dA.
\]
where $dA$ is the surface element at radius $r$, $\rho$ the gas density and $v_r$ the radial component of the velocity.

This operational definition measures radial gas circulation in the simulation and does not necessarily coincide with observationally inferred inflow or outflow rates. In observations, the interpretation depends on the tracer phase, emissivity weighting, projection effects, assumed geometry, and line of sight; we discuss these caveats in \S\ref{sec:disc}. Since a direct one‑to‑one mapping is not possible, throughout this section we discuss how these ideal inflows/outflows manifest in simulated data.

\subsubsection{Radial accretion properties}
Figure~\ref{fig:accr_radial} shows the time-averaged radial inflow profile (median lines with 1 sigma uncertainty shaded areas) for the two reference runs, together with the \stormy and \cloudy decomposition of the \high case. At the sink radius, for comparison, the circles report the time median sink accretion rate as measured in Figure~\ref{fig:bhar} with 1 sigma error bars (calculated over the elapsed time of the simulations). Outside the central kpc, all curves approach similar values and rise with radius following a power law trend (roughly $\propto r^{1.5}$), indicating that the large-scale atmosphere (within 100 kpc) provides a comparable mass supply in both turbulence regimes and it is only marginally affected by the presence of the jet as shown by their comparable slopes at the same radii in B26b. The main differences emerge inside the inner kpc. The \low run maintains a systematically larger inward mass flux down to the sink scale, consistent with its denser and more persistent central cold reservoir and with its prolonged super-Bondi accretion history. By contrast, the \high run shows a much weaker inflow in the central $\lesssim0.1$~kpc, despite comparable supply at larger radii. This implies that the reduction in late evolution of the BH feeding in the \high atmosphere is not caused by a lack of gas at meso- or macro-scales, but by a reduced efficiency in funnelling the condensed component through the inner transport bottleneck (Fig.~\ref{fig:f_couple}). 
At the sink micro-scale, the median SMBH accretion rates are \(\sim4\times10^{-3}\,M_\odot\,{\rm yr}^{-1}\) for \high and \(\sim0.15\,M_\odot\,{\rm yr}^{-1}\) for \low, illustrating the large difference in final coupling to the SMBH.

The \stormy/\cloudy decomposition of the \high run, shown in the dash-dotted and densely dotted lines, makes this point particularly clear. During the \stormy phase, the radial inflow profile is enhanced with respect to the \cloudy phase throughout the inner kpc, showing that precipitation is more directly coupled to SMBH feeding. In the \cloudy phase, instead, the inflow rate is strongly suppressed at the smallest radii, while remaining comparable at larger radii (see also C26a). This again supports the interpretation introduced above: in the \cloudy state, cold clouds and filaments still form and survive on micro- and meso-scales, but a smaller fraction of them is able to penetrate all the way to the sink.

To quantify the coupling between large-scale inflow and SMBH feeding, we define a scale-dependent coupling efficiency:
\begin{equation}
    f_{\rm couple}(r) \equiv \frac{\dot{M}_{\bullet}}{\dot{M}_{\rm in}(r)},
\end{equation}
where $\dot{M}_{\rm in}(r)$ is the mass inflow rate measured at radius $r$. This quantity provides a direct measure of how efficiently gas crossing a given scale is able to reach the SMBH. In Figure~\ref{fig:f_couple}, we compute $f_{\rm couple}$ as a function of the radii and, since the flow is highly time-variable and non-steady, this quantity is interpreted statistically through percentile-based measures. The coupling efficiency declines with radius in both runs, indicating that only a fraction of the inflowing gas at large scales is able to reach the SMBH. This reflects the inherently non-steady and multiphase nature of the flow, where mixing, stirring, and phase transitions progressively reduce the net inward mass transport. A clear systematic offset is observed between the two turbulence regimes. At all radii beyond the innermost region ($r>r_{\rm sink}$), the \low run exhibits a significantly higher coupling efficiency than the \high case. This indicates that, although both atmospheres are able to condense multiphase gas, only the \low configuration maintains an efficient inward transport toward the SMBH. 

The largest divergence occurs at meso-scales ($\sim 0.1-1$ kpc), where the \high run shows a rapid drop in $f_{\rm couple}$ by more than an order of magnitude relative to the \low case. This demonstrates that strong turbulence driving combined with the jet-induced turbulence disrupts the coherent inward cascade, effectively decoupling condensation from SMBH feeding. In terms of the weather framework, this behaviour suggests that, during the transition from a \stormy to a \cloudy state, cold gas remains abundant at meso-scales in the \high run, it is increasingly mixed, redistributed, or trapped in circulation, and only a small fraction is able to reach the central sink. By contrast, the \low run retains a more coherent, \rainy-like configuration, in which the condensed gas remains dynamically coupled to the SMBH.

The decomposition of the \high run into \stormy and \cloudy phases clarifies the physical origin of this behaviour. During the \stormy phase, the coupling efficiency remains relatively high across meso-scales, indicating that extended precipitation is still dynamically connected to the SMBH. In this regime, the larger jet-injected turbulence enhances condensation and promotes intermittent but effective inward transport, consistent with a bursty feeding. In contrast, the \cloudy phase is characterized by a sharp decline in $f_{\rm couple}$ at the meso-scale ($r\gtrsim300$ pc), with values suppressed by up to 2 orders of magnitude with respect to the \stormy phase. This demonstrates that, although cold gas remains abundant, it is no longer efficiently funneled to the centre. 
At the micro-scale, \(r\lesssim0.1\) kpc, the \stormy and \cloudy curves are more similar because the lower \(\dot M_\bullet\) in the cloudy phase is partly compensated by the reduced inflow reaching the same radii.

We interpret this transition as a shift from a transport-dominated to a mixing-dominated regime. In the \stormy state, filamentary structures and clouds retain sufficient coherence to sustain inward transport. In the \cloudy state, the cold phase is redistributed into a population of drifting clouds embedded in a disturbed hot halo. As a result, the multiphase medium becomes dynamically decoupled from SMBH feeding.

These results highlight that the presence of multiphase gas alone is not sufficient to ensure efficient accretion. The key quantity is the transport efficiency across the meso-scale, which is strongly regulated by the turbulent state of the atmosphere. In this framework, the \stormy-to-\cloudy transition marks the point at which the feedback-regulated system shifts from a precipitation-fed regime to a decoupled multiphase state, where cold gas survives but is no longer effectively accreted by the SMBH.

\subsubsection{Time-dependent multiphase transport across scales}
The phase-separated fluxes provide a complementary view. Figure~\ref{fig:inflow} displays the inflow rates calculated at different radii representative of the 3 spatial regimes, i.e. micro- (solid, 0.1 kpc), meso- (dashed, 1 kpc) and macro-scales (dotted, 10 kpc) as labelled in the legend. 
In observational terms, these ideal inflow/outflow patterns would likely translate into broader, multi-component ionized or molecular profiles in \stormy{} phases, and narrower, more coherent cold-gas kinematics in \rainy{} phases, although the exact mapping depends on projection and emissivity weighting.
On the other hand, large-scale hot outflows are visible primarily as X-ray cavities and shocks, rather than clear line-of-sight (LOS) velocity features in the optical or in molecular lines.

In the \low run (bottom), the total inflow increases with radius and is dominated at large scales by the hot phases, especially the hard- and soft-X-ray components. This indicates that the outer halo is largely governed by hot, volume-filling circulation. Once condensation is established (\(t/t_{\rm rain}\gtrsim1\)), however, the molecular-temperature, cold, and warm inflow rates at 0.1 and 1~kpc remain sustained (\(\gtrsim0.1\,M_{\odot}\,{\rm yr}^{-1}\)) for several rain times. Thus, while the total mass circulation is largest on macro-scales, the condensed component remains efficiently connected from the meso-scale down to the micro-scale, naturally supporting the persistent SMBH feeding typical of the rainy state. This behaviour matches the longer-lived central cold reservoir described in C26a, also typically observed in millimeter emission \citep[e.g.][]{ElfordDavis2024},  and naturally explains the persistent super-Bondi \bhar discussed above, typical of the \rainy weather where the gas feeding the innermost region is progressively transformed from a predominantly hot steady inflow at macro-scales into a multiphase, cold-dominated inflow at micro- and meso-scales. 

By contrast, in the \high run shown in the top graph, the same phases behave more episodically. During the \stormy weather, the molecular and cold inflow rates at 0.1~kpc rise rapidly and become comparable to those in the \low ($\sim0.1 \ M_{\odot} yr^{-1}$), indicating a vigorous precipitation directly feeding the SMBH. However, after $t/t_{\rm rain}\sim3.5$, the inner inflow in the condensed phases collapses by 2-3 orders of magnitude, while significant cold and molecular inflow can still be present at meso- and macro-scales. It is important to note that the \cloudy phase is not completely lacking inflowing condensed gas, but the coupling between meso- and micro-scale precipitation, eventually feeding the SMBH, becomes much weaker. This decoupling corroborates what was suggested by the accretion history in Figure~\ref{fig:bhar} and the thermodynamic phase distributions shown in C26a.

Outflow rates are shown in Figure~\ref{fig:outflow}. At \(r=10\)~kpc, the outward flux is dominated by the hot phases in both runs, with the hard-X-ray-temperature component providing the largest contribution. This large-scale hot outflow should be interpreted mainly as part of the volume-filling turbulent weather circulation of the halo, further stirred and reorganized by the jet, rather than as purely jet-ejected material. By contrast, the cool phases are more intermittent and primarily concentrated at smaller radii, consistent with uplifted/entrained condensates and local fountain motions rather than a coherent large-scale escaping cold wind.

In the \high\ case, the transition from \stormy\ to \cloudy\ is particularly evident. During the \stormy\ phase, up to \(t/t_{\rm rain}\sim4\), the outflow rates show flicker-like variability across phases, with the molecular-temperature, cold, and warm components becoming increasingly prominent as cooling proceeds. In the innermost shell (\(r=0.1\)~kpc), the cool outflow peaks around \(t/t_{\rm rain}\sim3\), reaching \(\sim0.1\,M_{\odot}\,{\rm yr}^{-1}\). After \(t/t_{\rm rain}\gtrsim4\), the inner outflow weakens, broadly following the decline in \bhar\ and the associated reduction in the central jet response. At the same time, cool outward transport becomes more radially extended: molecular-temperature, cold, and warm gas remain visible at the 1~kpc shell, and during the late \cloudy\ phase (\(t/t_{\rm rain}\gtrsim5\)) cold and warm material can intermittently reach the 10~kpc shell. The cold component can approach rates of order unity to a few \(M_{\odot}\,{\rm yr}^{-1}\), while the warm and molecular-temperature components remain lower and more intermittent. This late stage also shows larger relative variability, of order \(\sim1\) dex and even higher in the molecular-temperature phase, consistent with a more disturbed and turbulent multiphase medium (see also \S\ref{subsec:spectrum}).

By contrast, the \low\ run shows a more locally confined recycling cycle. The molecular-temperature, cold, and warm outflows are strongest within the inner \(\lesssim1\)~kpc, with typical rates spanning \(\sim10^{-5}-0.1\,M_{\odot}\,{\rm yr}^{-1}\). Weak and intermittent cool outward motions can appear at 10~kpc, but they are much less persistent and less massive than in the \high\ run. 
The soft-X-ray-temperature component also moves outward mainly on micro- and meso-scales, reaching the outer radial shell only at later stages (\(t/t_{\rm rain}\gtrsim6\)), qualitatively echoing recent X-ray studies of type-2 quasars in which extended soft-X-ray emission traces ionized gas shaped by AGN-driven outflows and multiphase feedback \citep[e.g.][]{TrindadeFalcaoKraemer2026}.
The hard-X-ray-temperature component, instead, is present at all epochs and radii, as expected for the hot phase that carries most of the volume-filling turbulent circulation. At 0.1 and 1~kpc, its outflow rate is more directly coupled to the central feedback response, since the injected jet material has \(T_{\rm jet}=5\times10^8\,{\rm K}\) and the injected mass flux scales with the SMBH accretion rate (Eq.~\ref{eq:mdot_out}). At 10~kpc, however, the hot outflow is smoother and remains around \(\sim10\,M_{\odot}\,{\rm yr}^{-1}\), suggesting that jet propagation, turbulent weather circulation, and halo-scale mixing wash out much of the short-timescale variability injected near the centre.

Overall, the mass-flux diagnostics reinforce the interpretation emerging from the thermodynamic and accretion analyses. The large-scale hot supply and volume-filling circulation are broadly comparable in the two runs, but the weather state determines how efficiently condensed gas is transported from meso- to micro-scales. In the \low\ run, and during the \stormy\ phase of \high, the condensed phases remain dynamically connected to the central sink and can sustain efficient SMBH feeding. In the late-time \cloudy\ phase of \high, instead, multiphase gas is still present on meso- and inner macro-scales, but a larger fraction is recycled, uplifted, or dispersed through the inner halo rather than being funneled into the SMBH.

\subsection{Kinematics – CCA diagnostics}\label{subsec:kin}
We now focus on two complementary diagnostics rooted in the CCA framework of \citet{Gaspari2018}: the k-plot, which traces projected multiphase gas kinematics, and the $\mathcal{C}$-ratio, which links cooling and turbulent mixing. Together, they provide observationally motivated probes of how the simulated weather states regulate meso-scale condensation and feeding.

\begin{figure*}
    \centering
    \includegraphics[width=0.85\linewidth]{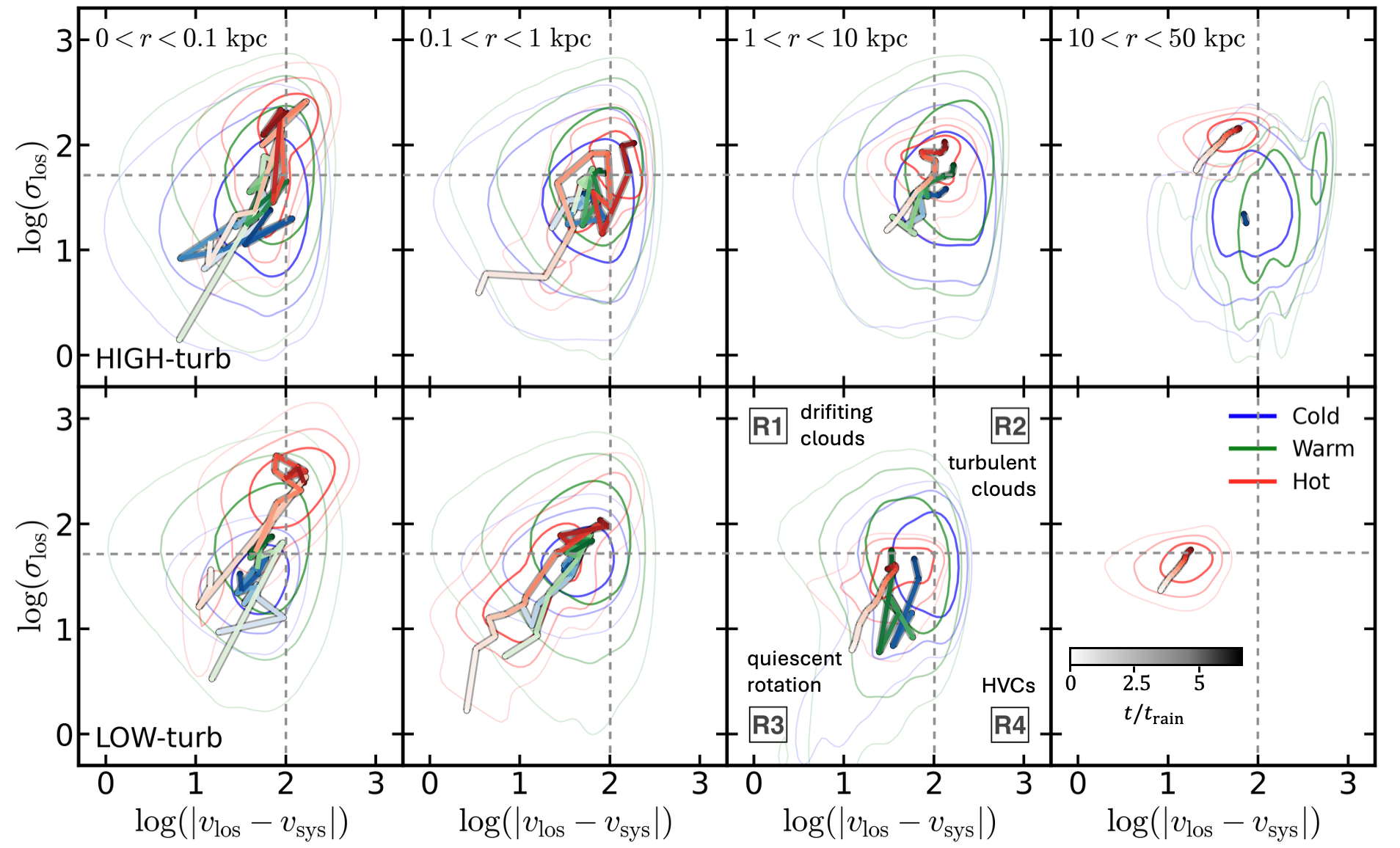}
    \caption{k-plot: evolution of projected gas kinematics in the k-plot plane, \(\log_{10}(|v_{\rm los}-v_{\rm sys}|/{\rm km\,s^{-1}})\) versus \(\log_{10}(\sigma_{\rm los}/{\rm km\,s^{-1}})\), shown for four radial shells from micro- to outer macro-scale. The top row shows \high, while the bottom row shows \low. Blue, green, and red contours show the time-integrated distributions of the cold, warm, and hot phase components, respectively. For visual clarity, the molecular-temperature and cold phases are grouped as ``cold'', while the soft- and hard-X-ray-temperature phases are grouped as ``hot''. Contours enclose the 85th, 92nd, and 97th percentiles of each phase distribution. Coloured tracks show the temporal evolution of the median position of each phase in the diagram, with darker colours corresponding to later times over the interval \(t/t_{\rm rain}=0-7\). Grey dashed lines mark the reference thresholds \(|v_{\rm los}-v_{\rm sys}|=100\,{\rm km\,s^{-1}}\) and \(\sigma_{\rm los}=50\,{\rm km\,s^{-1}}\), which separate the heuristic kinematic regions discussed in the text.}
    \label{fig:kplot_evo}
\end{figure*}

\begin{figure*}
    \centering
    \includegraphics[width=0.8\linewidth]{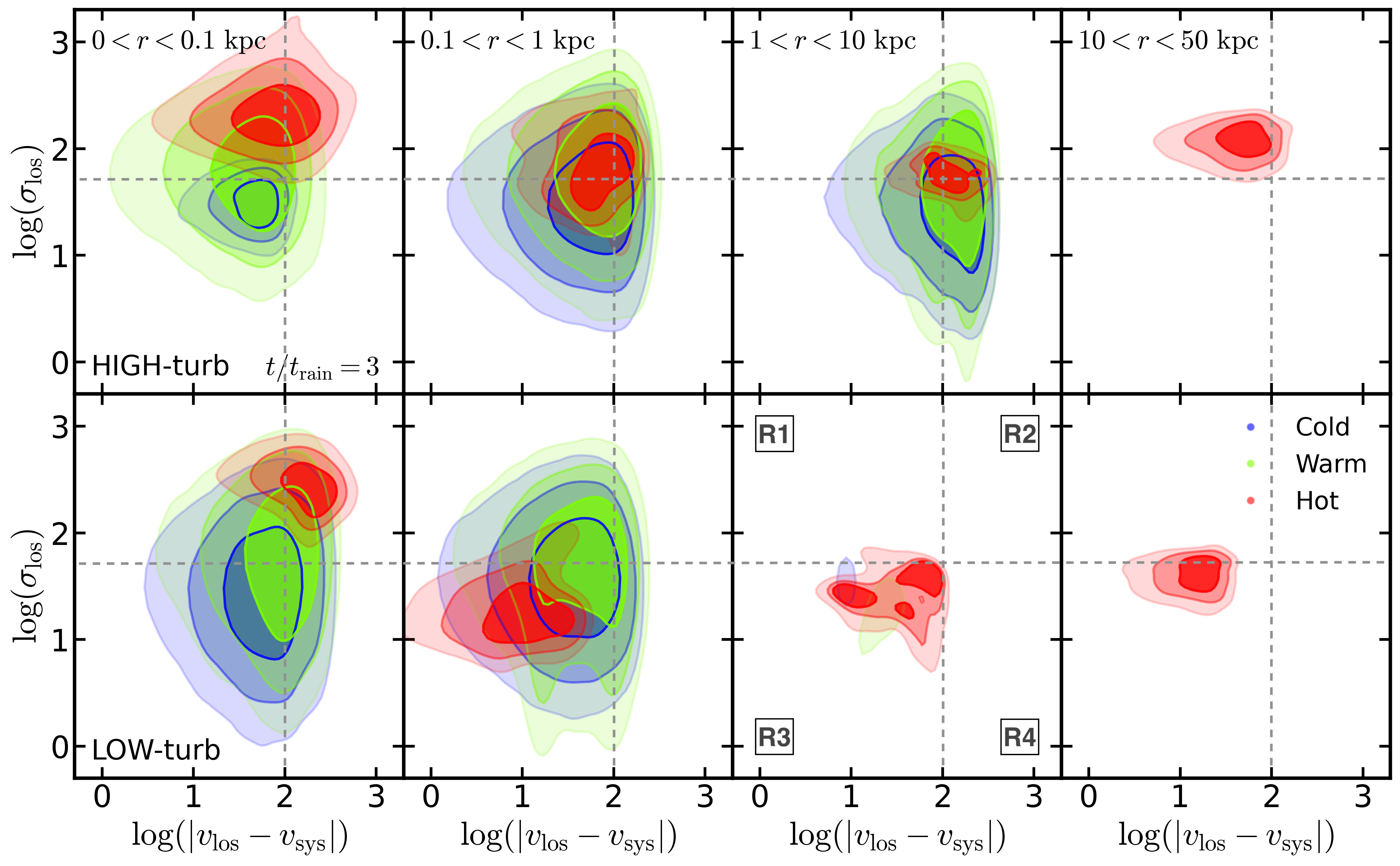}
    \includegraphics[width=0.8\linewidth]{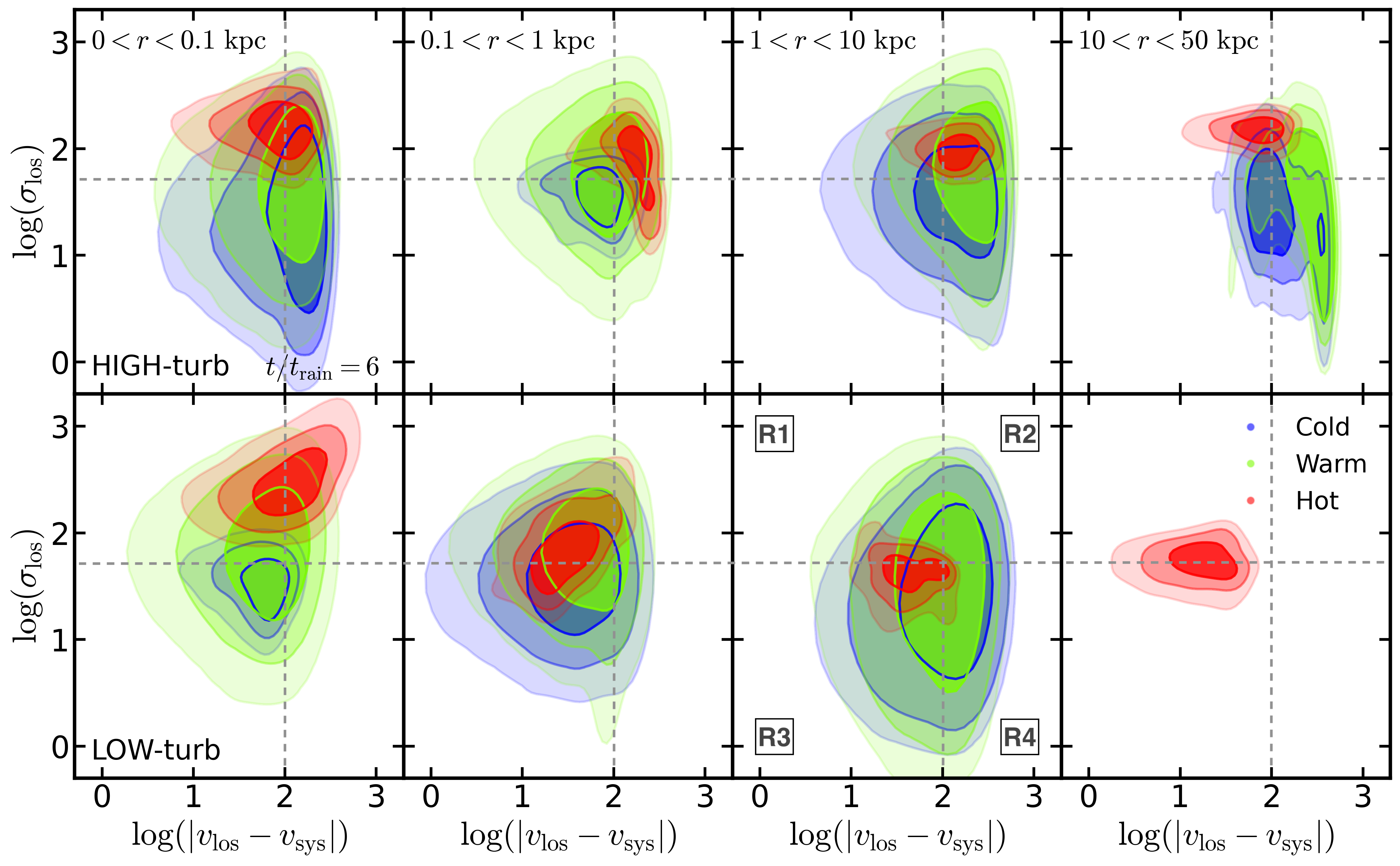}
    \caption{As in Fig.~\ref{fig:kplot_evo}, k-plots of the projected gas in radial bins, from the micro to the outer macro scale, for the \high and \low reference runs at two representative evolutionary stages. The upper block shows \(t/t_{\rm rain}=3\), while the lower block shows \(t/t_{\rm rain}=6\); within each block, the \high run is shown in the upper row and the \low run in the lower row. As in Fig.~\ref{fig:kplot_evo}, the molecular-temperature and cold phases are grouped as ``cold'', while the soft- and hard-X-ray phases are grouped as ``hot''. Solid contours and filled regions show the distribution of each phase in the corresponding radial bin, with hot, warm, and cold gas shown in red, green, and blue, respectively. The filled levels enclose the 85th, 92nd, and 97th percentiles, from lighter to darker shading.}
    \label{fig:kplot_frames}
\end{figure*}

\subsubsection{k-plots}\label{subsec:kplot}
To characterize the kinematics of the multiphase atmosphere, we construct the \textit{kinematical plot} (k-plot) following \citet{Gaspari2018}, in which we plot the line-of-sight velocity offset, $\log_{10}\!\left|v_{\mathrm{los}} - v_{\mathrm{sys}}\right|$, versus the line-of-sight velocity dispersion, $\log_{10}(\sigma)$, 
for beam-based projected samples \citep[to properly mock observational datasets, see also][]{Maccagni2021}. Specifically, we set $v_{\mathrm{sys}}=0$ km/s (we assume on average a null global bulk velocity of the gas with respect to the origin) and compute mass-weighted values, projected along the $z$ axis (hence the jet axis), for any given gas phase in separate radial ranges according to the definition of micro-, meso-, inner and outer macro-scale. It is important to note that the projected beam size is retrieved by considering a resolution buffer of 256 elements per side, where the projected area is adjusted to each radial shell (i.e., $l_{\rm side}=2\times r_{\rm shell,max}$).  Although the k-plot does not uniquely separate inflow from outflow, it provides an observationally motivated framework for classifying quiescent, rotating, drifting, turbulent, and high-velocity gas components in velocity-dispersion space.

As in B26b, we use the gray dashed lines as heuristic guides and divide the k-plot into four projected kinematic regions: drifting clouds (R1), with modest bulk velocity offsets but enhanced line broadening; strongly turbulent gas (R2), with both large velocity offsets and high dispersion; quasi-quiescent or rotation-dominated gas (R3), with low offsets and low dispersion; and high-velocity clouds (R4), with large projected bulk motion but comparatively lower internal dispersion. 
The kinematic regimes are defined by $|v_{\mathrm{los}}-v_{\mathrm{sys}}|=100~\mathrm{km \ s^{-1}}$ and $\sigma_{\mathrm{los}}=50~\mathrm{km \ s^{-1}}$. This representation is particularly useful because different regions of the diagram map naturally onto distinct cloud kinematics: the R3 quadrant corresponds to relatively quiescent clouds, the R2 area to turbulent clouds with both large velocity offsets and large internal dispersion, the R1 region to drifting clouds with modest bulk motion but large dispersion, and the R4 section to high-velocity clouds (HVCs) with stronger bulk motion but lower internal turbulence, as also observed in absorption-line studies \citep[e.g.][]{Tremblay2016}. The mutual interplay among the different phases, in terms of overlap and extent in the k-plot, is representative of the physical state of the gas. In particular, during CCA episodes, relatively quiet material is expected to progressively drift towards a chaotic state, with typical velocities of $50-100 \ \rm{km \ s^{-1}}$ \citep{Gaspari2018}. 

Figure~\ref{fig:kplot_evo} shows how the projected gas kinematics evolve in the k-plot for the two reference runs, separated into four radial shells and three broad thermal components. For visual clarity, we group the molecular-temperature and cold phases into a single cold component, and the soft- and hard-X-ray phases into a single hot component. The percentile contours show the time-integrated distribution of each phase, while the tracks follow the temporal evolution of the median kinematic state from early to late times. This representation allows us to track whether each phase remains kinematically ordered, becomes increasingly turbulent and mixed, or shifts toward regions of larger projected bulk motion. In this way, the k-plot links the simulated weather states to observable multiphase kinematic signatures.

A first common feature of both runs is the radial ordering of the kinematics. In general, all phases at most radii broadly follow a diagonal transition from the quiescent quadrant (R3) to the strong turbulence quadrant (R2). In the inner shells, all phases tend to occupy regions of relatively large velocity dispersion (from $10~\mathrm{km \ s^{-1}}$ up to $\sim500~\mathrm{km \ s^{-1}}$ in the hot phases), while at the same time the intrinsic velocities increase from a few $\times 10~\mathrm{km \ s^{-1}}$ up to $\gtrsim100~\mathrm{km \ s^{-1}}$ reflecting the stronger coupling between turbulence, jet forcing, and multiphase condensation near the centre. Moving outward, the characteristic loci become more compact, and the phase trajectories shorten, indicating that the kinematics becomes progressively less variable and more dominated by the global weather circulation. The outermost macro-scale shell ($10<r<50$~kpc) is the most stable in both runs and is largely set by the hot component, consistent with the fact that this region remains thermodynamically dominated by the diffuse hot X-ray atmosphere.

The main difference lies in the degree of kinematic mixing between phases in the inner kpc (see the first two columns). In the \high run, the trajectories are broader, more irregular, and more strongly overlapping. The cold and warm phases do not remain confined to a narrow locus; instead, they wander substantially in both \(|v_{\rm los}-v_{\rm sys}|\) and \(\sigma_{\rm los}\), often approaching the kinematic region occupied by the hot component. As a result, the condensed gas randomly samples a wide range of dynamical states, from relatively quiescent clouds (R3) to turbulent clouds (R2), and from high-velocity components (R4) to drifting clouds (R1). This is the kinematic counterpart of the \stormy multiphase atmosphere identified in C26a: stronger turbulence and a more disturbed jet cocoon continuously stir the condensed gas, enhance its kinematic overlap with the hot phase, and broaden the distribution of cloud velocities and dispersions. The hot phase also evolves rapidly from a more quiescent locus toward a turbulence-dominated regime, following the diagonal drift described above. Thus, in the \high run, the atmosphere is not only thermodynamically more mixed, but also kinematically more entangled.

The \low run instead displays a more coherent evolution. The phase trajectories are more compact and, particularly in the inner two shells, the cold, warm, and hot components remain more clearly separated through time. Their evolution is also smoother, with less pronounced wandering across the velocity-dispersion plane. This behaviour is consistent with the calmer weather state inferred from the morphological and phase-space diagnostics: condensation still occurs and the phases remain dynamically active, but the overall flow is less chaotic and the multiphase gas remains organized enough to sustain a persistent inward reservoir. The kinematic separation between phases is therefore better preserved in \low than in \high, indicating a more coherent coupling between precipitation and SMBH feeding.

The difference is especially informative in the inner kpc, where the simulations connect meso-scale precipitation to the central sink. In the \high case, the time trajectories of the cold and warm phases span a larger range in \(\sigma_{\rm los}\), with the colored contours covering \(\gtrsim 1\) dex at the 97th percentile, and remain closer to the hot component for a substantial fraction of the evolution. This indicates stronger stirring and repeated recycling between inflow, uplift, and local fountain motions. In the \low case, the same phases occupy tighter loci, with the cold component typically remaining within \(\lesssim 100~\mathrm{km\,s^{-1}}\) and spanning \(\lesssim 0.5\) dex at the 97th percentile. Their evolution also shows smaller excursions through the k-plane, consistent with a more coherent coupling between condensation and inward transport. Thus, the k-plot supports the interpretation already suggested by the \bhar, PSD, and inflow diagnostics: the \high run is characterized by a more chaotic, extended, and mixed multiphase circulation, consistent with the \stormy-to-\cloudy weather transition, whereas the \low run supports a more ordered and persistent condensation cycle, limited to the inner few kpc and consistent with a \rainy CCA regime.


Figure~\ref{fig:kplot_evo} summarizes the time-integrated distributions and median evolutionary tracks, while Figure~\ref{fig:kplot_frames} shows instantaneous k-plots at \(t/t_{\rm rain}=3\) and 6. The upper and lower blocks correspond to these two epochs, respectively; within each block, the \high and \low runs are shown in the first and second rows. As before, columns show the four radial shells, and colors indicate the cold, warm, and hot phase components. The contours trace the instantaneous phase distributions, with increasing opacity marking the 85th, 92nd, and 97th percentiles. This view highlights how the \stormy, \cloudy, and \rainy weather states populate projected velocity-dispersion space.

At \(t/t_{\rm rain}=3\), the \high case occupies a broader and more asymmetric region of the kinematic plane, especially in the inner shell. The cold and warm phases are shifted toward the R1 region, while the hot phase shows systematically larger \(\sigma_{\mathrm{los}}\) (\(\gtrsim100~\mathrm{km\,s^{-1}}\)) and \(|v_{\rm los}-v_{\rm sys}|\). This suggests that, at the micro-scale, the condensed gas is no longer confined to a narrow quiescent locus, but is being stirred and partially separated kinematically from the hot phase, consistent with the strong central accretion episode shown in Fig.~\ref{fig:bhar}. At meso- and inner macro-scales, the three phase components remain more dynamically mixed, indicating that the disturbed state extends beyond the immediate sink region. This is the kinematic imprint of the \stormy phase: the multiphase gas is strongly stirred, broadly mixed, and frequently displaced toward the turbulent-cloud and drifting-cloud regimes.
By contrast, the \low run at the same epoch retains a larger fraction of its cold and warm gas in the lower-left region of the plane (R3) at micro- and meso-scales. This is consistent with a population of more quiescent clouds that can remain coherently connected to inward precipitation and therefore contribute to the sustained \bhar. At the micro-scale, the hot component still shows large \(\sigma_{\mathrm{los}}\) and \(|v_{\rm los}-v_{\rm sys}|\) values, both \(\gtrsim100~\mathrm{km\,s^{-1}}\), but it shifts rapidly toward the R3 region at larger radii. The \rainy phase is therefore characterized by a more compact and coherent kinematic evolution, in which multiphase structures build up gradually and remain better organized across scales. In this sense, the high-turbulence atmosphere is already in a \stormy-dominated regime by \(t/t_{\rm rain}=3\), whereas the low-turbulence atmosphere is still developing toward a full \rainy weather state, consistent with the projections shown in C26a.

By \(t/t_{\rm rain}=6\), the difference becomes even more diagnostic. In the \high run, the micro-scale cold and warm phases have moved away from the earlier \stormy configuration and now occupy a broader region of the k-plot, extending mainly toward the R2 and R4 quadrants. Their \(\sigma_{\mathrm{los}}\) distribution spans up to \(\sim2\) dex around \(\sim50~\mathrm{km\,s^{-1}}\), while \(|v_{\rm los}-v_{\rm sys}|\) reaches \(\sim200{-}300~\mathrm{km\,s^{-1}}\). This indicates that a substantial fraction of the condensed gas is now more dispersed, turbulent, and characterized by large projected velocity offsets. Rather than forming a coherent inward reservoir, the cold and warm material is increasingly stirred and recycled, consistent with the weaker central accretion and reduced inner inflow shown in Figs.~\ref{fig:bhar} and \ref{fig:accr_radial}.
At the same time, the hot locus changes only mildly with respect to \(t/t_{\rm rain}=3\), suggesting that the hot phase has already reached a strongly stirred, quasi-steady kinematic state by this stage. At meso-scales, the cold and warm phases remain broadly similar in centroid position to their earlier configuration, although the contour area is reduced, consistent with less efficient late-time production and/or survival of centrally connected multiphase gas. At macro-scales, cold and warm gas have reached the outer shell 
with relatively large \(|v_{\rm los}-v_{\rm sys}|\sim100{-}300~{\rm km\,s^{-1}}\) and modest-to-intermediate dispersions of a few tens to \(\sim100~{\rm km\,s^{-1}}\), in agreement with C26a, where late-time cold filaments extend to \(\sim10{-}20\) kpc. Overall, this supports the picture developed from the accretion and inflow diagnostics: at \(t/t_{\rm rain}=6\), clouds and filaments are still present, but their coupling to direct SMBH feeding is weaker, and a larger fraction of the condensed gas participates in a turbulent, cloudy inner-halo circulation rather than in coherent inward transport.

The \low run instead follows a more stable and coherent evolution. Between \(t/t_{\rm rain}=3\) and \(6\), the micro-scale phases show only modest changes, with the cold component becoming more compact and less mixed with the warmer and hotter gas while remaining relatively quiescent. This behaviour is consistent with the sustained accretion rate \(\bhar\), since the condensed gas remains organized into a central reservoir rather than being dispersed into a broader turbulent circulation. At the meso-scale, the hot phase gradually shifts along the diagonal toward larger velocity dispersions and velocity offsets, reflecting the continued action of feedback and ambient turbulence. The cold and warm phases, however, move only mildly in the k-plot, indicating that the multiphase gas largely preserves its coherent kinematic structure at this scale.
This points to a gradual transition from quiescent clouds to moderately stirred condensates, as expected for a \rainy weather state. Because the spatial growth of the rain cycle is slower and more ordered than in the \high run, cold and warm components appear at the inner macro-scale only at later times, where they show broader scatter in both \(\sigma_{\mathrm{los}}\) and \(|v_{\rm los}-v_{\rm sys}|\). Even there, the phase loci remain more compact and show smaller excursions through the plane than in the \high case, indicating a more coherent kinematic evolution. Overall, the \low run is consistent with a calmer weather cycle in which condensed gas is progressively stirred while remaining comparatively well coupled to inward transport.

An interesting point of comparison can be drawn against the recent simulations shown in B26a,b within the same framework, where no-jet runs investigate the impact of turbulence driving and radiative cooling alone on CCA. Commonly to both studies, we see that the \stormy phase is usually well represented by a broad and mixed multiphase gas population across the \high k-plane while the \rainy weather is associated with more compact and kinematically distinct loci consistent with a more coherent dynamics. However, the k-plots shown in B26b (Figure~9) depict a situation where the phase distribution and the stochastic motion of gas condensates are partially slimmed down with respect to this work. This suggests that, despite the large-scale turbulent stirring provides sufficient kinematic support to sustain a multiphase CCA state, the jet-driven recycling and uplift added here are able to further amplify and favor multiphase mixing and gas circulation, resulting in broader and more overlapping contours highlighted in the k-plot diagnostic.  

The two kinematic snapshots reinforce the broader interpretation of the simulations. The \high run undergoes a genuine weather transition, from a \stormy phase dominated by strongly turbulent condensates to a \cloudy phase in which cold and warm gas remain present but become more dispersed, drifting, and weakly connected to direct SMBH feeding. The \low run does not show such an abrupt change of state; instead, it evolves more smoothly from quiescent clouds toward a moderately turbulent but still comparatively ordered multiphase flow. The k-plot therefore provides an observationally motivated diagnostic for distinguishing different modes of jet-regulated CCA: \stormy systems should occupy broader and more mixed regions of the velocity-dispersion plane, \cloudy systems should show extended but weakly coupled drifting condensates, and \rainy systems should retain more compact and coherent cold/warm loci.

\subsubsection{$\mathcal{C}$-ratios}\label{subsec:c-ratio}

\begin{figure*}
    \centering
    \includegraphics[width=0.85\linewidth]{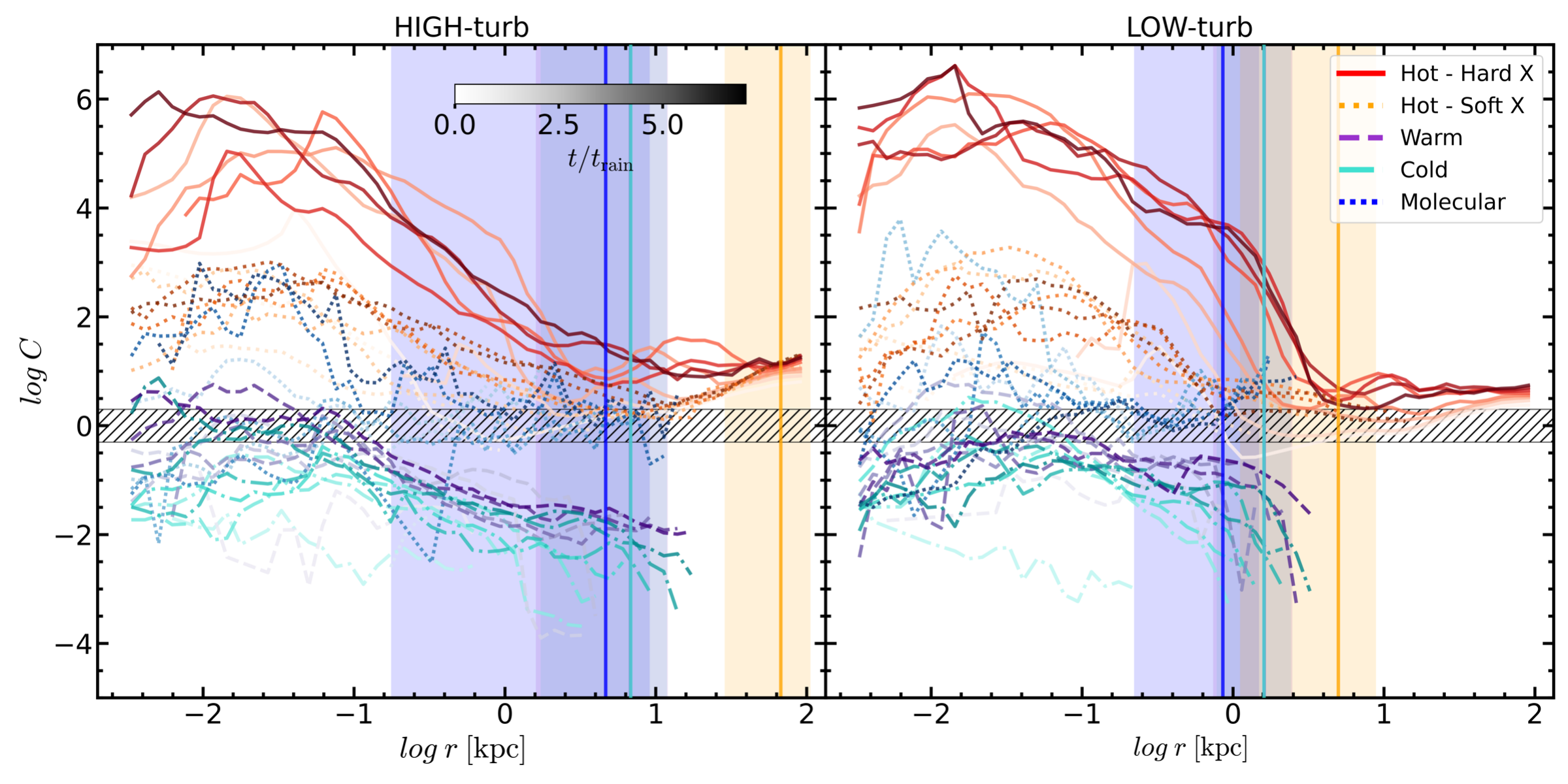}
    \includegraphics[width=0.85\linewidth]{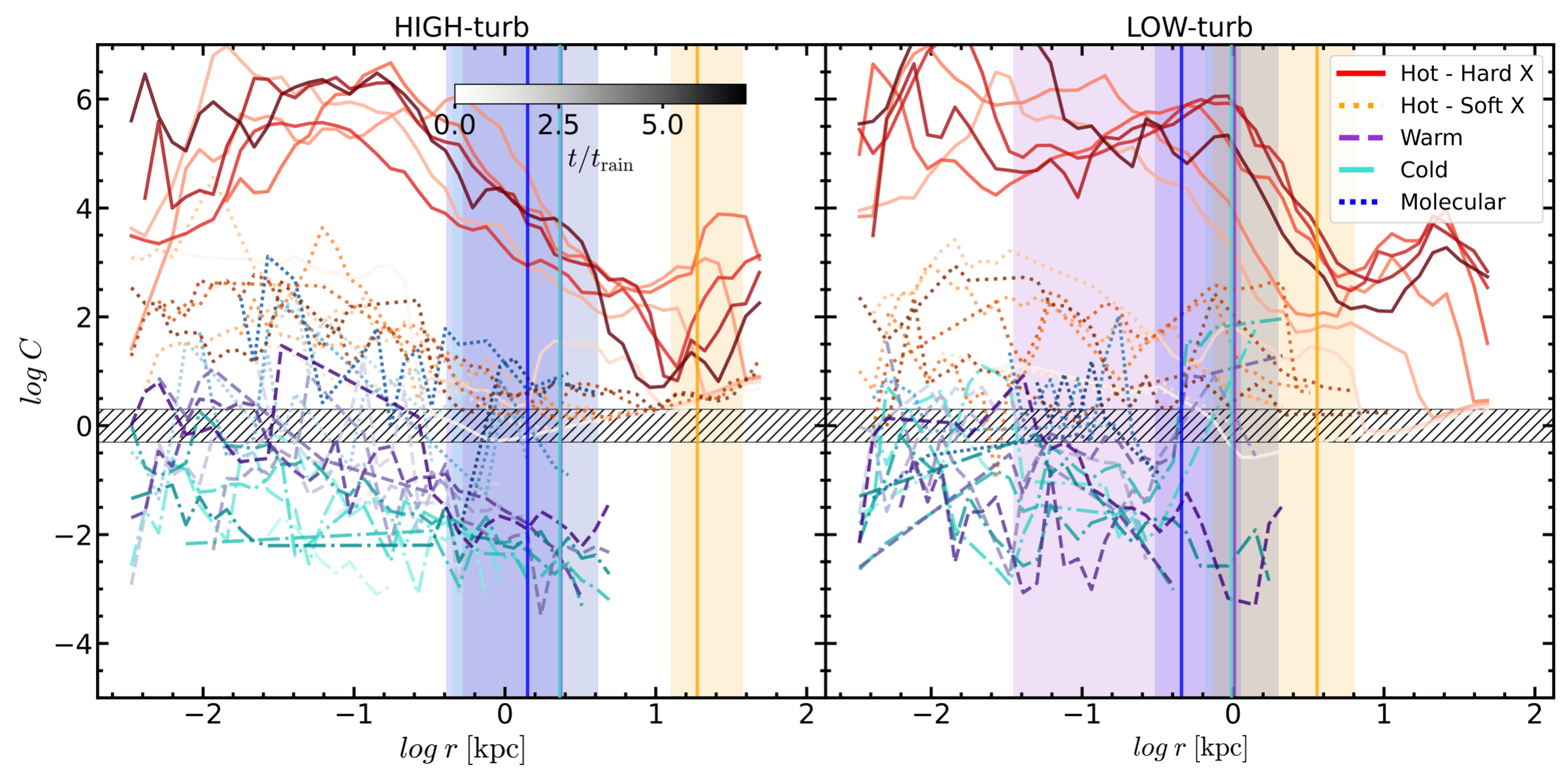}
    \caption{Radial evolution of the \cratio-ratio for different temperature-defined emission-band proxies, comparing the \high\ (left panels) and \low\ (right panels) runs. Curves with the same color and line style refer to the same gas phase, as indicated in the legend. Within each phase, lighter to darker shades encode increasing evolutionary time in units of \(t/t_{\rm rain}\), as shown by the color bar. Vertical lines and shaded bands mark the mean maximum radial extent of each phase and its scatter over the analyzed snapshots. The grey shaded region, $0.5<\mathcal{C}<2$, indicates the approximate condensation-prone regime where \(t_{\rm cool}\sim t_{\rm eddy}\). Upper panels show gas outside the bipolar jet cone, defined by a \(15^\circ\) half-opening angle around the fixed jet axis, while lower panels show gas inside the cone. The hard-X-ray phase reaches the largest central \cratio-ratio values and shows the steepest radial decline. The soft-X-ray and warm phases most clearly trace the approach to the condensation-prone band, while the cold and molecular-temperature phases should be interpreted as already-condensed gas tracked relative to the hot turbulent cascade.
    For the cooler phases, the plotted \cratio-ratio should be interpreted relative to the hot-phase turbulent cascade, rather than as a direct measure of the internal eddy time of individual clouds or filaments.}
    \label{fig:cratio}
\end{figure*}

We finally consider the ratio between the local cooling time and the eddy gyration time, typically referred to as $\mathcal{C}$-ratio. 
This quantity compares the timescale on which gas can radiatively lose thermal energy with the timescale on which turbulence creates/mixes gas perturbations.
Following \citet{Gaspari2018}, the dimensionless condensation ratio is
\begin{equation}
\mathcal{C} \equiv \frac{t_{\rm cool}}{t_{\rm eddy}}
\end{equation}
where the cooling time can be approximated as 
\begin{equation}
t_{\rm cool} \simeq \frac{3 k_B T}{n_e\,\Lambda},
\end{equation}
with $T$ being the gas temperature, $n_e$ the electron density, and $\Lambda(T, Z)$ the emissivity function.
The eddy gyration/turnover time quantifies how quickly subsonic turbulence stirs the gas on a given scale, and it can be estimated via
\begin{equation}
t_{\rm eddy} = 2\pi \,\frac{r^{2/3}\,L^{1/3}}{\sigma_{v,L}},
\end{equation}
where $L$ is the turbulence injection scale and $\sigma_{v,L}$ is the velocity dispersion measured at that scale. This expression assumes a Kolmogorov cascade, $\sigma_v(\ell)\propto \ell^{1/3}$ \citep{kolmogorov1941}, as found for subsonic turbulence in simulations \citep[e.g.][]{GaspariChurazov2013,Fournier2025}. At radius $r$, the characteristic eddy size is expected to be of order $r$ (larger eddies cannot be contained locally), hence $\ell\sim r$.\footnote{Given the large dynamic SMR range, estimating the turbulent amplitude from \(\sigma_v(r)\) (capped at the driving scale $L$) measured in spherical radial shells and \(L\sim r\) is a reasonable numerical approximation.}

This quantity provides a direct measure of the propensity of the atmosphere to undergo turbulence-regulated condensation. Large values, \cratio$\gg1$, indicate that the cooling time is much longer than the eddy turnover time. In this regime, turbulent stirring and mixing act faster than radiative losses, efficiently dispersing overdense perturbations before they can cool nonlinearly. By contrast, \cratio$\ll1$ implies that cooling proceeds much faster than turbulent mixing, so the flow tends toward a more coherent cooling-flow regime rather than a turbulence-regulated CCA cascade. The CCA regime is therefore expected to occur when \cratio approaches unity, within the typical $\sim0.3$ dex scatter around unity \citep[i.e. $t_{\rm cool}\sim t_{\rm eddy}$; see also B26b,][]{Gaspari2018}. In this regime, turbulence can seed nonlinear density perturbations while radiative cooling amplifies them into multiphase clouds and filaments. Such condensation can persist even in the presence of AGN feedback, provided the jet does not fully erase the local balance between cooling and turbulent mixing, as suggested by observations of galaxies, groups, and clusters across different spectral bands \citep[e.g.][]{Juranova2020, Maccagni2021, Temi2022, OlivaresSalome2022, Singha2023, Mehdipour2023, Lepore2025}.

We split the volume into two regions using a bipolar cone of half-opening angle $15^\circ$ centered on the fixed jet axis. This cone is a diagnostic mask, not the numerical injection geometry: the injected jet remains collimated along $\pm z$ without an imposed opening angle. It is therefore not meant to track jet-launched or entrained material directly, but only to separate the jet-channel/interface region from the off-axis ambient atmosphere. We stress that, in the present analysis, the \cratio-ratio is computed for all temperature phases using the eddy turnover time inferred from the hot turbulent cascade, \(t_{\rm eddy,hot}\). This choice is physically motivated because the \cratio-ratio criterion is fundamentally a hot-phase diagnostic: it identifies where turbulence in the volume-filling atmosphere allows radiative cooling to amplify perturbations into multiphase condensation. It should therefore not be interpreted as a direct measurement of the internal eddy time of already-condensed filaments or clouds. For the cooler phases, the plotted \cratio-ratio values should instead be regarded as phase-tracking diagnostics relative to the ambient hot turbulent field.\footnote{
A fully phase-dependent estimate would require measuring the characteristic size and internal velocity dispersion of individual cold clouds and filaments. Since these structures become progressively smaller and more fragmented at lower temperatures, using the large-scale hot-phase eddy time may underestimate the intrinsic \cratio-ratio of the cooler gas. A clump-by-clump analysis of this kind is beyond the scope of the present work and will be investigated in a future study.}

Figure~\ref{fig:cratio} compares the radial profiles of the \cratio-ratio across emission-band proxies for two reference runs, \high (left panels) and \low (right panels), while the outside/inside jet-cone regions are depicted in the upper/lower figures, respectively. Different line styles and colors indicate the various gas phases, while the color intensity traces the progression of the simulation in time as a function of $t/t_{\rm rain}$. Vertical lines and related shaded areas show the average spatial extent in radius of each phase with their 1 sigma uncertainty over the range of snapshots considered in the plot.
Figure~\ref{fig:cratio} can be read in three complementary ways. First, the vertical position of each curve distinguishes mixing-dominated gas, \cratio$\gg1$, condensation-prone gas, \cratio$\sim1$, and rapidly cooling gas, \cratio$\ll1$.\footnote{For the already-condensed cold and molecular-temperature phases, however, values below \cratio$\sim1$ should not be read as a separate cooling-flow criterion, but as the location of post-condensation gas relative to the hot-phase eddy time.} 
Second, the radial interval near \cratio$\sim1$ identifies where CCA rain can be sustained. Third, the inside/outside cone comparison traces feedback anisotropy: the hot jet channel remains at high \cratio-ratio values, while the surrounding atmosphere more readily approaches the CCA threshold.

We start by analyzing the \cratio-ratio outside the jet cone, shown in the upper panels. Similarly to B26a, the hard-X-ray phase in the \high\ run shows relatively low values, \cratio$\sim1-10$, at radii larger than a few kpc, while the \cratio-ratio rises steeply toward the centre, reaching \cratio$\sim10^{4}-10^{6}$ at \(r<1\) kpc. This radial trend is mainly driven by the rapid decrease of \(t_{\rm eddy}\) toward small radii, whereas the cooling time of the hard-X-ray gas varies more mildly. The hard-X-ray phase therefore approaches the condensation-prone band, \cratio$\simeq0.5-2$, primarily at \(r\sim1-10\) kpc during the early stages of the run, \(t/t_{\rm rain}\sim1-2\). This radial interval marks the region where the hot atmosphere is most susceptible to extended nonlinear condensation, consistent with the broad multiphase fragmentation characteristic of the \stormy\ weather state identified in C26a.
At later times, the remaining hard-X-ray gas is displaced toward larger \cratio-ratio values, \cratio$\gtrsim10$, indicating that turbulent stirring and mixing dominate over radiative cooling in this phase. The soft-X-ray phase, by contrast, lies systematically closer to the condensation-prone regime, with \cratio-ratio values lower by about \(2-3\) dex, especially at micro- and meso-scales. This suggests that the soft-X-ray-temperature gas acts as the intermediate cooling stage between the volume-filling hot atmosphere and the warmer and colder condensates. 
Its proximity to \cratio$\sim1$ over a broad radial range marks the soft-X-ray phase as the main thermodynamic gateway between the volume-filling hot atmosphere and the warmer and colder condensates.
Consistently, the maximum radial extensions shown by the vertical bands reach a few tens of kpc for the soft-X-ray phase and \(\lesssim10\) kpc for the warm, cold, and molecular-temperature components.

In the \low\ run, the hard-X-ray phase follows a qualitatively similar radial behaviour, but the transition from the outer, flatter profile to the inner rising profile is steeper, and the \cratio-ratio remains closer to unity at \(r\gtrsim5\) kpc. 
This indicates that condensation-prone conditions are reached more coherently, although over a more compact radial region, as traced by the smaller radial extent of the soft-X-ray, warm, cold, and molecular-temperature bands. Thus, reaching \cratio$\sim1$ in the hot phase is a necessary indicator of precipitation-prone gas, but the radial extent of the colder phases also depends on the amplitude of perturbations and the ability of condensates to survive and remain connected across scales.
This behaviour is consistent with B26b, where no jet was included, suggesting that outside the directly affected jet cone the large-scale CCA condition is still primarily regulated by the ambient turbulent cascade and cooling structure.
Following the condensation cascade toward cooler phases, we find that the average radial extent of the warm, cold, and molecular-temperature components is more confined than in the \high\ run. In particular, the cold and molecular-temperature phases remain mostly within the inner kpc, supporting the interpretation of a compact and coherent \rainy\ weather state, as also found in B26b. The molecular-temperature \cratio-ratio profiles provide a particularly useful imprint of this weather transition. In the late-time \high\ run, the molecular-temperature profiles show a marked increase toward the centre, especially below \(0.1\) kpc. This is driven by the lower central cold-gas density in the \cloudy\ phase relative to the earlier \stormy\ phase. For approximately fixed temperature and cooling function, \(t_{\rm cool}\propto1/n\), so a density decrease lengthens the cooling time and increases the \cratio-ratio. By contrast, the \low\ run does not show this central rise. Instead, it evolves toward higher molecular-temperature gas densities, keeping the inner \cratio-ratio closer to \cratio$\sim1$ and favoring a persistent, reservoir-fed \rainy\ inflow.

By contrast, inside the jet-cone, shown in the lower panels, the hard-X-ray phase remains at systematically high \cratio-ratio values, \cratio$\sim10^{2}-10^{6}$, in both the \high\ and \low\ runs, especially at small radii \(r\lesssim1\) kpc. This indicates that the jet channel is comparatively hostile to sustained in-situ condensation. The kinetic outflow excavates a hot, low-density, dynamically stirred region in which turbulent mixing and jet heating dominate over local radiative cooling. The gas outside the cone, less directly affected by the injected jet power, is therefore more likely to approach the multiphase rain threshold. The warm and cold phases inside the cone nevertheless show broader temporal variability, including intermittent excursions toward the condensation-prone regime, \cratio$\sim1$. These episodes likely trace transient cooling, uplift, shocks, and mixing layers driven by jet-induced instabilities. They may seed local and short-lived condensation events, but they do not indicate that the jet channel is the dominant site of long-lived precipitation. Finally, inside the cone the \high\ and \low\ runs show no strong systematic difference within the intrinsic scatter of the \cratio-ratio profiles. This suggests that, close to the jet axis, jet-driven stirring and heating dominate over the imposed ambient turbulent driving, reducing the contrast between the two turbulence amplitudes.

One robust result is the strong geometrical dependence of the condensation criterion. Outside the jet cone, the warm, cold, and molecular-temperature phases remain closer to the condensation-prone regime, \cratio$\sim1$, over an extended radial interval, especially from the meso-scale to the inner macro-scale. This indicates that sustained precipitation is favoured primarily in the ambient stirred atmosphere and jet-interface regions, rather than along the jet channel itself, as confirmed also in the simulations with the inclusion of the SMBH spin (P26b). This picture is consistent with the multiphase morphology and inflow diagnostics, where cold gas can form and circulate outside the directly heated jet cone before part of it is transported inward. This differs from the pure feeding simulations of B26b, where the cold and molecular phases remain much more cooling-dominated, with typical molecular values \(C\sim10^{-2}\)--\(10^{-3}\). The comparison suggests that jet uplift and feedback-driven mixing/reheating keep part of the condensed gas closer to the hot turbulent atmosphere.
For the hot phases, the outside-cone behaviour broadly agrees with the no-jet results of B26a,b, suggesting that the large-scale cooling-versus-turbulence balance remains primarily controlled by the ambient turbulent cascade. The cooler phases, however, show \cratio-ratio values larger by about \(\sim1\) dex, especially at \(t/t_{\rm rain}\gtrsim5\). This likely reflects the additional stirring and velocity dispersion induced by the jet as its energy couples to the surrounding atmosphere. Thus, even outside the nominal jet cone, feedback can broaden the turbulent cascade and make the condensation-prone regime more intermittent, particularly in the late-time cloudy state. 

The inside/outside comparison is particularly informative for the \high\ case. During the \stormy\ stage, the region outside the cone is driven rapidly toward the condensation-prone regime, \cratio$\sim1$, over a substantial radial interval, matching the epoch of strongest precipitation and SMBH accretion. At later times, as the system enters the \cloudy\ phase, the relevant phases still remain close to the condensation band outside the cone, but in a less coherent and more temporally variable way. This indicates that clouds can continue to form, while their coupling to direct SMBH feeding becomes weaker. Conversely, inside the cone the hot phases remain displaced toward high \cratio-ratio values, confirming that the jet channel itself is not the primary site of sustained condensation.

Overall, the two turbulence regimes differ mainly in how broadly, coherently, and persistently they approach the condensation-prone regime. In the \high\ run, the \cratio-ratio profiles show larger temporal excursions and a broader radial spread, especially outside the cone and for the warm and cold phases. This reflects the stormier atmospheric conditions already identified through the kinematic and accretion diagnostics: stronger stirring broadens the region over which gas can be driven toward \cratio$\sim1$, but also makes that condition more intermittent and irregular. The \low\ run instead displays a smoother and more coherent evolution. Outside the cone, the cooler phases approach \cratio$\sim1$ more steadily and over a more compact radial range, while inside the cone the separation between the hot and condensed gas remains cleaner. This behaviour is consistent with a calmer \rainy\ weather state, in which condensation is less explosive but more persistently maintained.

Taken together, the \cratio-ratio diagnostics reinforce the main physical picture emerging from C26a and B26a,b. Multiphase condensation is favoured where turbulent stirring and radiative cooling become comparable, \(t_{\rm cool}\sim t_{\rm eddy}\), and this condition is reached most naturally in the ambient atmosphere surrounding the jet rather than within the jet channel itself. The main jet-regulated effect is therefore geometrical: feedback displaces sustained precipitation away from the hot jet channel and into the surrounding turbulent atmosphere and interface layers. Outside the cone, the \high\ run reaches the condensation-prone regime over a broader and more intermittent radial interval, consistent with an extended \stormy\ precipitation cycle followed by a weaker \cloudy\ coupling phase. The \low\ run approaches the same regime more coherently and over a more compact region, supporting a persistent, reservoir-fed \rainy\ state. Inside the cone, cold and warm gas can still show intermittent excursions toward \cratio$\sim1$, likely associated with jet-driven mixing layers, shocks, and local instabilities. These episodes should be interpreted as transient interface events, not as evidence that the jet channel is the dominant site of long-lived CCA rain.

\section{Discussion}\label{sec:disc}

This work assesses how the kinematic and temporal diagnostics of CCA are modified when radiative cooling and driven turbulence are coupled to an actively responding kinetic jet. Together with the companion thermodynamic analysis presented in C26a, our results show that the jet does not simply oppose cooling through local heating, nor merely advect pre-existing condensates. Instead, it dynamically reorganizes the multiphase medium, changing where gas condenses, how it moves across scales, and how efficiently it remains connected to SMBH feeding.

The two turbulence regimes, \high and \low, differ primarily in their inner transport efficiency. Outside the central few hundred parsecs, both runs maintain substantial mass circulation, showing that the halo continues to supply gas to the inner atmosphere. The decisive difference appears closer to the sink. In the \low case, condensed gas remains well coupled from meso- to micro-scales and sustains a long-lived, super-Bondi accretion state. In the \high case, late-time inner inflow is strongly suppressed despite the continued presence of multiphase gas at larger radii. Strong ambient turbulence therefore does not shut off condensation altogether; it weakens the final coupling between condensation and SMBH feeding.

This behaviour maps naturally onto the weather states introduced in C26a. The early \high evolution corresponds to a \stormy phase, with spatially extended precipitation, strong phase mixing, bursty super-Bondi accretion, and rapidly multiphase inflow/outflow channels. At later times, the same run evolves toward a \cloudy state: clouds and filaments persist across meso- and inner macro-scales, but are less coherently connected to inward transport. The accretion-rate PSD flattens at low frequencies, and the SMBH feeding rate drops by more than an order of magnitude relative to the preceding \stormy episode. The \low run instead maintains a smoother \rainy mode, in which the central cold reservoir survives longer and sustains the feedback cycle.

The variability, kinematic, and \cratio-ratio diagnostics reinforce this picture. Both runs show flicker-like low-frequency variability and steep high-frequency red-noise tails, consistent with a scale-coupled multiphase feeding cycle rather than stochastic white noise. However, the \low case retains more variability power over long intervals because its central reservoir remains connected to the SMBH, whereas the \high case transitions from a bursty \stormy episode to a weaker \cloudy regime with reduced long-timescale coherence. The k-plot provides the corresponding kinematic view: the \high run evolves from strongly turbulent condensates to more dispersed and drifting clouds, while the \low run evolves more monotonically from quiescent to moderately stirred multiphase gas. The \cratio-ratio profiles show that condensation-prone conditions are preferentially reached in the ambient atmosphere surrounding the jet channel, rather than inside the jet cone itself. In the \high run this regime spans a broader radial range, whereas in the \low run it remains more centrally concentrated. These diagnostics identify the meso-scale as the mediating layer of jet-regulated CCA, where turbulence, cooling, uplift, and phase conversion determine whether precipitation becomes extended and weakly coupled, or compact and efficiently feeding.

An instructive comparison is provided by P26a,b, where the jet direction is coupled to the evolving SMBH spin rather than kept fixed. Even modest reorientations, of order a few degrees, allow the jet-driven stirring to be distributed over a larger effective volume, making the transition toward a \sunny{} state more persistent and more evident than in the fixed-axis runs analyzed here. In P26b, this broader coupling more efficiently depletes cold gas from the central and meso-scale regions, producing a clearer \stormy{}-to-\sunny{} evolution. By contrast, in the fixed-axis \high{} run presented here, cold gas is never fully erased: cold inflow and outflow remain visible down to micro- and meso-scales even during the late \cloudy{} phase, although their coupling to SMBH feeding is strongly reduced. The radius--time maps in Appendix~\ref{app:rt_maps} and Fig.~\ref{fig:rt_maps} make this distinction explicit. A \cloudy{} state retains multiphase gas with inefficient central accretion, whereas the \sunny{} state reported in P26b corresponds to a more sustained clearing of the cold component from the inner transport region.

Several limitations should be taken into account. The present work analyzes two benchmark turbulence amplitudes and adopts a fixed-axis, purely kinetic, mass-loaded jet. The simulations neglect magnetic fields, anisotropic conduction and viscosity, cosmic rays, star formation, stellar feedback, detailed molecular chemistry, and relativistic jet propagation. For instance, allowing a reorientation of the jet axis would certainly introduce an extra turbulence and mixing component, especially where magnetic pressure and reconnection are expected to play a key role in shaping the gas environmental conditions, hence affecting multiphase condensation and CCA precipitation. 

A potential caveat of the present setup is the adoption of a purely kinetic jet, neglecting explicit thermal, magnetic, or cosmic-ray feedback channels at injection. In realistic AGN environments, the relative partition between these components is uncertain and likely varies with the accretion state and microphysical conditions. However, our choice should also be viewed as a controlled and physically motivated simplification. By injecting feedback purely in kinetic form, we avoid imposing an a priori subgrid prescription for thermal energy deposition, allowing thermalization to arise self-consistently through shocks, turbulent dissipation, and mixing in the resolved flow. This approach enables us to isolate the role of kinetic energy in regulating multiphase condensation and gas transport, and to assess how efficiently mechanical feedback alone can couple to the atmosphere across scales.

In addition, unresolved disk-to-horizon physics is captured through fixed subgrid efficiencies, and the k-plot diagnostics are not yet full emissivity-weighted mock observations including projection and instrumental effects. These choices isolate how ambient turbulence modifies jet-regulated CCA, but likely underestimate the diversity of coupling pathways in realistic systems.

\subsection{Comparison with observed AGN feeding and feedback}
\label{subsec:obs_comparison}

The diagnostics introduced here can be compared with multiphase observations of AGN host galaxies, groups, and clusters, although they are not yet forward-modeled observables. Our coldest bin should be interpreted as a molecular-temperature proxy, not as a direct prediction of CO luminosity or molecular mass, and the warm and X-ray phases are not emissivity-weighted mock H$\alpha$, [N\,{\sc ii}], or X-ray maps. With this caveat, the most useful comparisons are filament extent, projected velocity offsets, velocity dispersions, hot-gas turbulent amplitudes, accretion-state indicators, and condensation thresholds. Table~\ref{tab:obs_mapping} summarizes a proposed mapping between simulation diagnostics and observational proxies. It should be read as a semi-quantitative guide rather than a one-to-one forward model. A direct comparison with individual systems will require synthetic CO, optical-line, and X-ray diagnostics, including emissivity weighting, projection, instrumental resolution, and tracer selection; this will be investigated in future work.

A recurring result from ALMA, MUSE, and X-ray studies is that cold molecular and warm ionized filaments are spatially associated with X-ray cavities, rims, and uplifted low-entropy gas \citep[e.g.][]{Tremblay2016,Tremblay2018,TemiAmblard2018,VantyghemMcNamara2019,OlivaresSalome2019,OlivaresSalome2022,Olivares2023,Olivares2025}. Observed structures span compact central reservoirs of \(\lesssim3\) kpc in some group and early-type systems, while cool-core cluster filaments commonly extend over \(\sim3-25\) kpc. Their projected velocities are typically \(|v_{\rm los}-v_{\rm sys}|\sim100-400~{\rm km~s^{-1}}\), often below the free-fall expectation \citep{Tremblay2018,Russell2019,VantyghemMcNamara2019,OlivaresSalome2019}. This is broadly consistent with the jet-regulated CCA picture inferred here: the jet does not simply remove or heat the condensing gas; it excavates a hot channel and reorganizes the surrounding atmosphere, where cooling, entrainment, fallback, and feeding coexist.

In this context, the \high{} and \low{} runs map onto different parts of the observed multiphase phenomenology. The \high{} run resembles extended filamentary systems: during the \stormy{} and \cloudy{} phases, cold and warm gas reach the inner macro-scale and occupy kinematic loci with \(|v_{\rm los}-v_{\rm sys}|\sim100-300~{\rm km~s^{-1}}\) and moderate dispersions, similar to uplifted or slowly raining filaments. The \low{} run instead remains more centrally concentrated, with molecular-temperature and cold gas predominantly retained in the inner kpc and only a slower extension to larger radii.
Group-centred systems are especially relevant for the present setup. In NGC~5846, NGC~4636, and NGC~5044, ALMA CO(2-1) observations reveal off-centre molecular clouds with broad linewidths and irregular velocity structure, interpreted as unbound molecular associations drifting through the turbulent hot atmosphere rather than as a settled rotating disk \citep{TemiAmblard2018,Temi2026}. This provides a closer kinematic analogue to our cold-gas transport than a comparison based on molecular mass. 

Warm ionized gas provides a complementary kinematic comparison. Optical/IFU and multiphase studies of radio galaxies, brightest group/cluster galaxies, and nearby AGN hosts show complex velocity fields, moderate line-of-sight offsets, and substantial broadening, indicating that the gas is often neither a settled disk nor a purely radial wind \citep{Maccagni2021,OlivaresSalome2022,Singha2023, EskenasyOlivares2024, EskenasyOlivares2026}. This is the regime targeted by the k-plot. The thresholds used here, \(|v_{\rm los}-v_{\rm sys}|=100~{\rm km~s^{-1}}\) and \(\sigma_{\rm los}=50~{\rm km~s^{-1}}\), lie within the observed multiphase range \citep{Gaspari2018,Maccagni2021}. In \high{}, cold and warm gas repeatedly cross these thresholds, especially in the inner and meso-scale regions. In \low{}, the phase loci remain more compact. Thus, cold or warm gas near cavities or radio structures should not be classified from morphology alone as inflow or outflow, as it can often trace local CCA circulation.

Two observed cases are particularly useful anchors. In Fornax~A, MeerKAT, ALMA, and MUSE data reveal concurrent feeding and feedback, including turbulent clouds, a \(\sim3\) kpc filament, and a larger-scale multiphase outflow in the wake of the radio jets \citep{Maccagni2021}. This resembles the mixed inflow, uplift, and circulation seen in the \high{} run, although Fornax~A includes additional environmental and merger-related complexity. On smaller scales, molecular absorption in cool-core BCGs reveals cold gas with velocities between \(-45\) and \(283~{\rm km~s^{-1}}\) relative to systemic, likely biased toward motions in the direction of the SMBH \citep{Rose2019}. This is particularly relevant to the inner k-plot interpretation: the \low{} run and the early \high{} \stormy{} phase resemble efficiently coupled cold-cloud feeding, whereas the late \high{} \cloudy{} phase retains cold gas with weaker final coupling to the sink.

The hot X-ray phase sets the turbulent background for condensation. Cavities, shocks, rims, and low-entropy structures show that AGN feedback acts mainly through mechanical redistribution rather than spatially uniform heating \citep{Fabian2012,GittiBrighenti2012,McNamaraNulsen2012}. Direct calorimeter measurements now anchor the velocity scale across different hot-atmosphere states: Hitomi measured $\sigma_{\rm los}=164\pm10~{\rm km~s^{-1}}$ in Perseus, while XRISM found $\sigma_v=169\pm10~{\rm km~s^{-1}}$ in A2029, representative of calm subsonic cores \citep{Hitomi2016,Xrism2025}. At the disturbed end, XRISM observations of the merging cluster Abell 2034 show much broader lines, with $\sigma_v\simeq470~{\rm km~s^{-1}}$, a kinetic pressure fraction of $\sim15\%$, and a possible turbulent Mach number $\mathcal{M}\sim0.5$ \citep{HeinrichZhang2026}. Indirect constraints from X-ray density and thermodynamic perturbations likewise indicate predominantly subsonic motions, but with substantial system-to-system variation \citep{GaspariChurazov2013,Hofmann2016}. Thus, our runs should be compared with observations at the level of turbulence regime rather than as synthetic line profiles: \texttt{low-turb} ($\mathcal{M}_{\rm 3D}\sim0.15$) is close to calm Hitomi/XRISM-like cores, whereas \texttt{high-turb} ($\mathcal{M}_{\rm 3D}\sim0.4$) represents a more strongly stirred but still subsonic atmosphere, closer to disturbed XRISM systems such as Abell 2034.

The simulated accretion states are also broadly consistent with mechanically dominated AGN feedback. Both runs peak near \(\lambda_{\rm Edd}\simeq8\times10^{-5}\), span roughly \(\lambda_{\rm Edd}\sim10^{-6}-10^{-1}\), and spend a significant fraction of their evolution in a sub-Eddington regime. Observed Eddington ratios are inferred from bolometric luminosities, bolometric corrections, emission-line proxies, and black-hole mass estimates \citep{VestergaardPeterson2006,Lamastra2009,Duras2020}, so our accretion-rate-based \(\lambda_{\rm Edd}\) is only an analogue. Nevertheless, the simulated range is consistent with low-luminosity radio-mode systems and broad survey-inferred accretion-rate distributions \citep{KauffmannHeckman2009,Aird2012,BestHeckman2012,HeckmanBest2014}, with rare high-\(\lambda_{\rm Edd}\) excursions corresponding to short-lived CCA bursts rather than sustained quasar-like phases \citep{Gaspari2017,Sadowski2017}. Nuclear wind studies provide a complementary small-scale view of the same feedback cascade \citep{Mehdipour2023}.

The \cratio-ratio profiles provide the closest thermodynamic connection. In the CCA framework, nonlinear condensation is expected when radiative cooling and turbulent mixing operate on comparable timescales, namely \(C\equiv t_{\rm cool}/t_{\rm eddy}\approx1\). In \citet{Gaspari2018}, the observed condensation locus is centered on \(C\simeq1\), with multiphase systems falling within a band close to \(\pm0.3\) dex interval around unity, similar to our reference shaded interval. Observed filaments generally lie within low-entropy, short-cooling-time regions, often where \(t_{\rm cool}/t_{\rm ff}\lesssim20\), and in several systems near \(t_{\rm cool}/t_{\rm eddy}\sim1\) \citep{OlivaresSalome2019,OlivaresSalome2022,Lepore2025}. In our simulations, this \(C\sim1\) band is reached most systematically outside the jet cone. The \high{} run reaches it over a broader radial interval, consistent with extended \stormy{}/\cloudy{} filaments, whereas \low{} reaches it in a more compact central region, consistent with \rainy{} precipitation.

These comparisons highlight the main observational implication: multiphase gas alone is not sufficient evidence for efficient SMBH feeding. In the \high{} \cloudy{} state, cold and warm gas remain visible on meso- and inner macro-scales while the inner inflow and SMBH accretion rate decline. Systems with rich ALMA/MUSE filamentary structure but weak or intermittent instantaneous AGN activity may therefore occupy a weakly coupled \cloudy{} state, in which condensation continues but only a small fraction of the cold phase reaches the central engine.

\section{Conclusions}
\label{sec:concl}

In this work, we have analyzed the kinematics, radial inflow/outflow structure, accretion variability, and CCA diagnostics of a jet-regulated multiphase atmosphere in a galaxy-group environment. This paper complements C26a, which focuses on the morphology and thermodynamics of the same simulations. Within the {\sc BlackHoleWeather} framework, the setup builds on the CCA and self-regulated feedback picture developed in \citet{GaspariBrighenti2011,Gaspari2013,Gaspari2017,Gaspari2018} and on the no-jet turbulent CCA baseline of B26a,b. We examined two benchmark hydrodynamical simulations including radiative cooling, driven subsonic turbulence, pc-scale sink accretion, and a self-regulated kinetically-dominated jet. The two runs, \high{} and \low{}, differ only in the imposed turbulent driving strength, allowing us to isolate how ambient turbulence modifies jet-regulated feeding and feedback. Combined with C26a, our main conclusions are as follows.

\begin{enumerate}

    \item Ambient turbulence regulates both the onset and the geometry of jet-regulated CCA. In \high{}, condensation starts later but develops into a broader, more filamentary, and more strongly mixed multiphase structure. In \low{}, condensation starts earlier and remains more coherent, centrally concentrated, and persistent.

    \item The main difference between the two weather regimes is inner transport efficiency, not the large-scale gas supply. Both runs maintain substantial inflow and circulation at meso- and macro-scales, but only \low{} preserves strong coupling from meso- to micro-scales. In the late \high{} evolution, multiphase gas remains abundant outside the centre, while the final inward transport to the sink is weakened.

    \item Once precipitation begins, both runs become genuinely CCA-fed. The accretion rate rises from Bondi-like values to strongly super-Bondi levels ($\dot{M}_\bullet/\dot{M}_{\rm Bondi}\sim10^{2-3}$), showing that smooth hot-mode accretion is replaced by multiphase, precipitation-driven feeding. The \low{} run sustains this state for longer, whereas \high{} undergoes a short, intense \stormy{} burst followed by a weaker and more intermittent \cloudy{} mode. Both simulations remain mostly in a low-Eddington, mechanically dominated regime, with sporadic episodes at larger accretion states.

    \item Accretion variability retains the flickering character expected for CCA. The accretion-rate PSDs show flicker-like low-frequency slopes and steep red-noise high-frequency tails, consistent with scale-coupled multiphase feeding. Compared with the no-jet B26b baseline, the two jet-regulated runs show more similar break frequencies, suggesting that jet-driven stirring partly reduces the direct dependence of the variability timescale on the imposed ambient turbulence. In \high{}, the transition to the \cloudy{} phase lowers the PSD normalization and flattens the low-frequency slope, indicating that coherent, large-scale rain episodes are replaced by weaker and less correlated feeding events.

    \item The phase-separated mass fluxes show how the weather state controls cold-gas transport. In \high{}, molecular-temperature, cold, and warm gas become more radially extended and participate in uplift, fountain-like recycling, and cloudy circulation out to kpc and inner-macro scales. In \low{}, cool recycling remains mostly confined to the inner kpc, favoring a compact reservoir that feeds the SMBH more persistently.

    \item The k-plot and the \cratio-ratio profiles provide complementary diagnostics of whether cold gas is dynamically linked to SMBH feeding. The \cratio-ratio profile identifies where the hot atmosphere is thermodynamically prone to condensation, with \(C=t_{\rm cool}/t_{\rm eddy}\sim1\), and shows that this condition is reached most systematically outside the jet cone and around the jet-ambient interface. The k-plot shows whether the resulting cold and warm gas is coherent, turbulent, drifting, or dispersed. Neither diagnostic alone is sufficient: cold gas can form where condensation is allowed, yet still be dynamically decoupled from the sink. Together, the two diagnostics show that \high{} develops a broader condensation-prone region but more dispersed and weakly coupled cold gas, whereas \low{} maintains a more compact \(C\sim1\) region and more coherent cold/warm kinematics, favoring sustained feeding.

\end{enumerate}

Overall, jet feedback acts primarily as a dynamical reorganization mechanism rather than as simple local heating. The kinetic jet excavates a hot, low-density channel, suppresses sustained in-situ condensation inside the cone, and promotes uplift, entrainment, mixing, and circulation in the surrounding atmosphere and jet-ambient interface. The \textit{meso-scale} therefore does not merely host multiphase condensation: it sets the transport bottleneck between halo cooling and SMBH feeding. The presence of cold gas alone is insufficient to infer efficient accretion; what matters is whether the local weather state allows condensed gas to remain thermodynamically favoured, kinematically coherent, and dynamically connected to the inner inflow. In this sense, the {\sc BlackHoleWeather} framework \citep{Gaspari2020} provides the physical language and diagnostic structure needed to connect macro-scale halo cooling, meso-scale multiphase transport, and micro-scale SMBH feeding. This extends the B26a,b no-jet CCA results to the jet-regulated regime and motivates the combined use of k-plot and \cratio-ratio diagnostics to interpret multiphase AGN feeding and feedback.\\

\begin{acknowledgements}
The {\sc BlackHoleWeather} authors acknowledge key funding support from the European Research Council (ERC) under the European Union's Horizon Europe research and innovation programme (Consolidator Grant {\sc BlackHoleWeather}, No.~101086804; PI: Gaspari). Views and opinions expressed are, however, those of the author(s) only and do not necessarily reflect those of the European Union or the European Research Council Executive Agency; neither the European Union nor the granting authority can be held responsible for them. 
We acknowledge ISCRA for awarding this project access to the LEONARDO supercomputer, owned by the EuroHPC Joint Undertaking, hosted by CINECA (Italy).
We acknowledge EuroHPC JU for awarding this project access to JUPITER, hosted by the Jülich Supercomputing Centre (JSC), Germany, under project EHPC-REG-2025R02-366 and to EHPC-REG-2025R02-368.
We acknowledge EuroHPC Joint Undertaking for awarding us access to Karolina at IT4Innovations (Czech Republic) and MeluXina at LuxProvide (Luxembourg).
VO acknowledges support from the DICYT ESO-Chile Comite Mixto PS 1757, Fondecyt Regular 1251702, and CIRAS-AI Project, code FIUF137139-USACH.
PT acknowledges support from NASA NNH22ZDA001N Astrophysics Data and Analysis Program under award 24-ADAP24-0011.
RS acknowledges funding from the CAS-ANID grant No.~220016.
FMM carried out part of the research activities described in this paper with contribution of the Next Generation EU funds within the National Recovery and Resilience Plan (PNRR), Mission 4 - Education and Research, Component 2 - From Research to Business (M4C2), Investment Line 3.1 - Strengthening and creation of Research Infrastructures, Project IR0000034 – “STILES - Strengthening the Italian Leadership in ELT and SKA.
We thank Silvano Molendi, Francesco Tombesi, Martin Fournier, and Amirnezam Amiri for comments and discussions.
We thank the organizers and participants of the following conferences and workshops for the stimulating discussions that helped improve this work: `BlackHoleWeather I' (Sexten, ITA), `SMBH-2025' (G\"oteborg, SWE), and `Breaking the Limits 2026' (Cagliari, ITA).
\end{acknowledgements}

%
%

%
\bibliography{biblio/my}{}
\bibliographystyle{aa}


\onecolumn
\appendix

\section{Synthetic observational mapping}\label{obs_table}

\setcounter{table}{0}
\renewcommand{\thetable}{A.\arabic{table}}

\begin{table*}[!h]
\centering
\caption{
Synthetic mapping between simulation diagnostics and observational proxies.
The entries summarize how jet-regulated CCA weather states may appear in
multiphase observations. They are intended as an interpretive guide, not as
forward-modelled observables.
}
\label{tab:obs_mapping}
\small
\setlength{\tabcolsep}{4pt}
\renewcommand{\arraystretch}{1.25}
\begin{tabular}{p{0.18\linewidth} p{0.23\linewidth} p{0.29\linewidth} p{0.22\linewidth}}
\hline\hline
\textbf{Simulation diagnostic} &
\textbf{Observable proxy} &
\textbf{Observed anchor} &
\textbf{Weather-state cue} \\
\hline

\multicolumn{4}{l}{\emph{Multiphase morphology}} \\
\hline

Cold/warm extent and distribution &
CO, H$\alpha$, [N\,{\sc ii}] morphology; compact reservoirs; off-centre clouds &
Cold/warm gas ranges from compact nuclear reservoirs to kpc-scale filaments,
often associated with cavities, rims, or uplifted low-entropy gas.\tablefootmark{a} &
\high{}: extended \stormy{} or \cloudy{} filaments.
\low{}: compact \rainy{} reservoir. \\

\hline
\multicolumn{4}{l}{\emph{Kinematics and circulation}} \\
\hline

Projected cold/warm motions &
CO, H$\alpha$, [N\,{\sc ii}] centroids; absorption components &
Offsets of order \(10^2~{\rm km~s^{-1}}\) are common in cold/warm filaments and
nuclear absorbers, often below pure free-fall expectations.\tablefootmark{b} &
Large offsets with modest dispersion suggest uplift, fallback, or weakly coupled
\cloudy{} circulation. \\

Velocity dispersion / k-plot &
CO and optical line widths; kinematic phase-space location &
The adopted k-plot thresholds,
\(|v_{\rm los}-v_{\rm sys}|=100~{\rm km~s^{-1}}\) and
\(\sigma_{\rm los}=50~{\rm km~s^{-1}}\), sample the observed multiphase range.\tablefootmark{b} &
\stormy{}: broad/turbulent loci.
\rainy{}: compact coherent loci.
\cloudy{}: drifting or dispersed gas. \\

Inflow/outflow ambiguity &
Line asymmetry, centroid shifts, gas near bubbles and radio structures &
Cold/warm gas near cavities or jets can trace uplift, fallback, local fountains,
or feeding, rather than a single radial flow.\tablefootmark{b} &
CCA circulation: inflow, uplift, mixing, and fallback coexist. \\

\hline
\multicolumn{4}{l}{\emph{Hot atmosphere and condensation}} \\
\hline

Hot-phase stirring &
X-ray calorimeter line broadening; X-ray fluctuations; cavities and rims &
Direct spectroscopy and fluctuation analyses constrain predominantly subsonic
hot-atmosphere motions, even in disturbed systems.
\tablefootmark{c} &
\low{}: calmer atmosphere, compact precipitation.
\high{}: stronger stirring, broader circulation. \\

\(\cratio=t_{\rm cool}/t_{\rm eddy}\) &
Cooling time plus turbulence constraints &
CCA condensation is expected within an interval around unity \(C\sim0.5-2\).\tablefootmark{d} &
Extended \(C\sim1\): \stormy{}/\cloudy{} condensation.
Compact central \(C\sim1\): \rainy{} condensation. \\

\hline
\multicolumn{4}{l}{\emph{AGN state}} \\
\hline

Accretion level and variability &
Nuclear luminosity, cavity power, radio duty cycle, long-term monitoring &
Observed \(\lambda_{\rm Edd}\) estimates are luminosity-based; radio-mode systems
trace mechanically dominated, low-Eddington feedback with intermittent activity.\tablefootmark{e} &
\stormy{}: bursty feeding.
\rainy{}: sustained coupling.
\cloudy{}: weak instantaneous feeding. \\

\hline
\end{tabular}

\tablefoot{
\tablefoottext{a}{Multiphase morphology and filament extent:
\citet{Tremblay2016,Tremblay2018,TemiAmblard2018,OlivaresSalome2019,VantyghemMcNamara2019,OlivaresSalome2022,Olivares2025,Russell2019}.}
\tablefoottext{b}{Cold/warm kinematics, absorption, and k-plot diagnostics:
\citet{Gaspari2018,Tremblay2018,OlivaresSalome2019,Rose2019,Maccagni2021,Singha2023}.}
\tablefoottext{c}{Hot-atmosphere feedback and turbulence constraints:
\citet{Fabian2012,GittiBrighenti2012,McNamaraNulsen2012,GaspariChurazov2013,Hofmann2016,Juranova2020,Hitomi2016,Xrism2025,HeinrichZhang2026}.}
\tablefoottext{d}{CCA condensation thresholds and cooling/turbulence diagnostics (and hot/warm filament structure):
\citet{Gaspari2018,OlivaresSalome2019,OlivaresSalome2022,Olivares2025,Lepore2025}.}
\tablefoottext{e}{AGN accretion-state and mechanical-feedback context:
\citet{VestergaardPeterson2006,Lamastra2009,KauffmannHeckman2009,Aird2012,BestHeckman2012,HeckmanBest2014,Duras2020,Gaspari2017,Gaspari2020,Sadowski2017,Mehdipour2023}.}
}
\end{table*}

\section{Time-radius maps}\label{app:rt_maps}

\begin{figure*}
    \centering
    \includegraphics[width=0.85\linewidth]{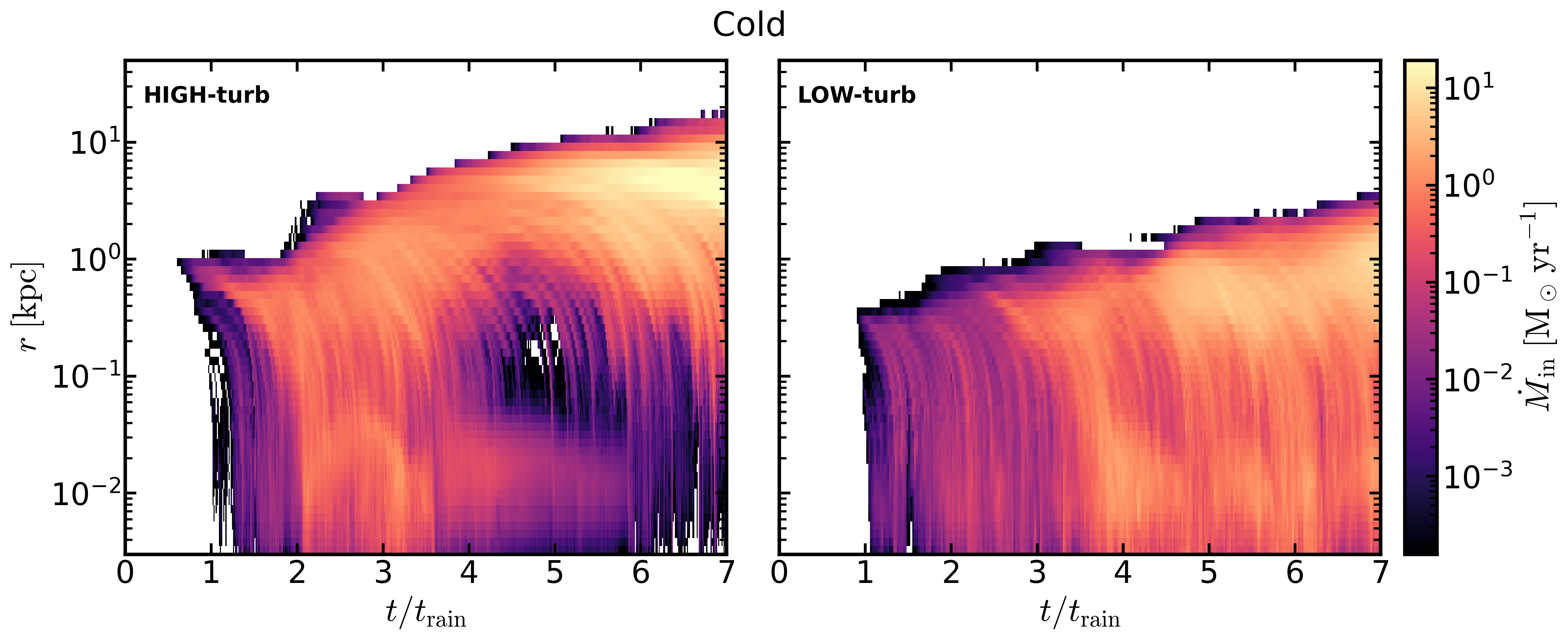}
    \includegraphics[width=0.85\linewidth]{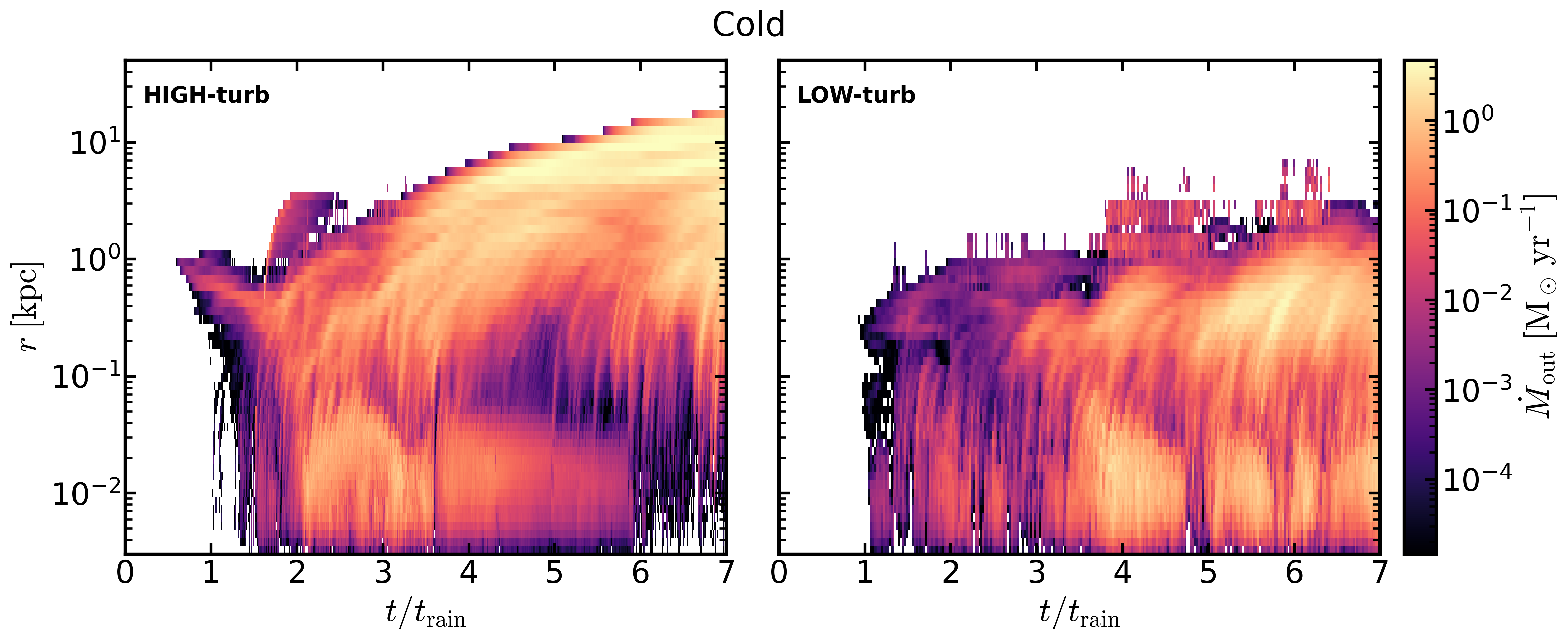}
    \caption{Radius--time 2D histograms of the inflow (upper panels) and outflow (lower panels) mass rates for the combined molecular-temperature and cold gas phase (\(T<1.6\times10^4\,{\rm K}\)). The \high{} run is shown in the left panels and the \low{} run in the right panels. Mass rates are computed as the total inflowing or outflowing mass flux in each radial shell, using 50 radial bins sampled every 1 Myr.}
    \label{fig:rt_maps}
\end{figure*}

To further support the weather-phase distinction, Fig.~\ref{fig:rt_maps} shows radius--time 2D histograms of the inflow (top panels) and outflow (bottom panels) mass rates, \(\dot{M}_{\rm in}\) and \(\dot{M}_{\rm out}\), for the combined molecular-temperature and cold gas phase (\(T<1.6\times10^{4}\,{\rm K}\)). The maps are computed as total mass fluxes in 50 radial shells over \(0<r<50\) kpc and sampled every 1 Myr, with the \high{} run shown on the left and the \low{} run on the right. They highlight the difference between the late \cloudy{} state in the fixed-axis \high{} run and the more prevalent \sunny{} clearing reported in P26b. In the present \cloudy{} state, cold gas is still present from micro- to inner-macro scales even at \(t/t_{\rm rain}\gtrsim6\), but its coupling to central accretion is inefficient. In the P26b \sunny{} state, by contrast, jet-axis reorientation more persistently suppresses the cold component across the inner transport region.

\end{document}